# 2020 NDSA AGENDA FOR DIGITAL STEWARDSHIP

An NDSA Publication

A report on the challenges, opportunities, gaps, emerging trends, and key areas for research and development that support the global capacity for digital stewardship.

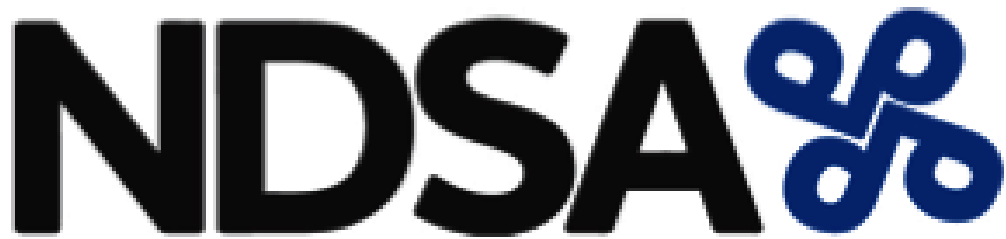

AUTHORS
Micah Altman, MIT Libraries, at the Massachusetts Institute of Technology
Karen Cariani, WGBH Media Library and Archives
Bradley Daigle, Academic Preservation Trust; University of Virginia Library.
Christie Moffatt, National Library of Medicine (NLM), National Institutes of Health
Sibyl Schaefer, University of California, San Diego
Bethany Scott, University of Houston Libraries' Special Collections
Lauren Work, University of Virginia
Representing the NDSA Agenda Working Group

## TABLE OF CONTENTS

## ABOUT THE NDSA

Founded in 2010, the NDSA is an international membership organization that supplies advocacy, expertise, and support for the preservation of digital heritage. The NDSA promotes a vision in which all digital material fundamentally important to our cultures receives appropriate, effective, and sustainable stewardship care from the international preservation community to protect and enhance its persistent value, availability, and (re)use. NDSA member institutions represent all sectors, and include universities, consortia, non-profits, professional associations, commercial enterprises, and government agencies at the federal, state, and local levels.

More information about the NDSA is available at http://www.ndsa.org.

# 1. Executive Summary

## 1.1 Where are we at the dawn of 2020? What could we be in 2025?

The 2015 Agenda set out to provide "funders, decision-makers, and practitioners with insight into emerging technological trends, gaps in digital stewardship capacity, and key areas for research and development to support the work needed to ensure that today's valuable digital content remains accessible, useful, and comprehensible in the future, supporting a thriving economy, a robust democracy, and a rich cultural heritage."[1] This edition of the NDSA Agenda reflects on the work in the digital stewardship community that has occurred since 2015 and frames the priorities that we believe should be the focus of the digital preservation community as a whole over the next five years. As part of the 2020 Agenda, a survey was conducted by the NDSA Agenda Working Group in 2017-2018 to solicit feedback and information about digital curation and preservation priorities from key decision makers at NDSA member organizations. The results of this survey, which helped to inform this 2020 Agenda, can be found in the Appendix.

While the results of the survey and evolving digital stewardship strategy have resulted in a completely updated *Agenda*, the report remains organized into the same four overarching topic areas as the 2015 *Agenda*—Building Digital Content Collections, Organizational Policies and Practices, Technical Infrastructure Development, and Research Priorities. Each section is briefly summarized below and provides a review of developments, an analysis of priorities, and a set of recommendations aimed at addressing the highest-priority challenges for the digital stewardship community.

A number of themes emerge across these topic areas. First, the preservation community is innovating: the report offers many new and important examples of materials that are now being safeguarded; community practices refined; innovative organizational initiatives launched; and greater understanding of new research offering deeper awareness of preservation threats and processes. Second, that community's progress is at times unsteady and the systematic, broad impact of these efforts remains challenging to measure: the registries that would enable straightforward determination of which content is actively protected and which is known to be at risk are too localized; there is a lack of comprehensive surveys of preservation practices and their effectiveness; and standardized methods and testbeds for conducting generalizable research are often wanting. Third, collective action is essential to success—the scale of content in need of preservation goes beyond the resources of individual institutions, and the economics of replication and use create opportunities for collective action to have broad benefits with modest investments. Finally, collective action is very hard, and the organizations that coordinate and scaffold collective action remain vulnerable—as demonstrated by the sunset of Digital Preservation Network,[2] D-Lib[3], and the near-demise of the Keeper's Registry.[4]

---

[1] National Digital Stewardship Alliance, "2015 National Agenda for Digital Stewardship," 2015, http://www.digitalpreservation.gov/documents/2015NationalAgenda.pdf.

### 1.1.1 Building Digital Content Collections

The 2015 *National Agenda* made four core recommendations related to Building Digital Content Collections: to support partnerships, donations, and agreements with creators and owners of digital content and stewards; to build the evidence base for evaluating at-risk, large-scale digital content for acquisition; to understand the technical implications of acquiring large-scale digital content; and to share information about what content is being collected and what level of access is provided.

**Where are we in 2020?**

Over the last several years, the digital stewardship community has made sustained progress in several initiatives around supporting partnerships with donors, creators, and other stewards of content. Work on the Documenting the Now (DocNow)[5] project, for example, has led the conversation on the value and importance of involving and building community around the collecting of digital content, particularly social media, in ethically responsible ways. Launched in 2016 with support from the Andrew W. Mellon Foundation, DocNow has hosted a number of meetings, workshops, and symposia, including the 2018 Ethics in Archiving the Web Symposium. Another great example of partnership is the collaboration between the Council on Library and Information Resources (CLIR)'s Digital Library Federation (DLF) and the Historically Black Colleges & Universities (HBCU) Library

---

[2] "Information Update," The Digital Preservation Network, December 5, 2018, https://web.archive.org/web/20190226161550/http://dpn.org/news/2018-12-05-information-update.

[3] "D-Lib Magazine," Corporation for National Research Initiatives, accessed April 16, 2020, http://www.dlib.org/.

[4] "The Keepers Registry Funding Cessation Banner Announcement," April 25, 2019, http://web.archive.org/web/20190425225022/https://thekeepers.org/.

[5] "DocNow," Documenting the Now Project, accessed April 8, 2020. https://www.docnow.io/

Alliance,[6] which is exploring the common ground between the two communities around digital libraries and digital library-based pedagogy. The 2017 report *Common Mission, Common Ground*[7] identified the forging of an authentic working partnership between communities and articulated significant conversations around inclusion, representation, recruitment and retention of underrepresented groups in libraries, and more inclusive digital library collections. This working partnership has now received an additional 3-year IMLS grant to fund 15 year-long fellows from HBCUs for mentoring and professional development beginning in 2019.[8]

Notable new efforts have emerged in the area of building the evidence base for evaluating large-scale digital content for acquisition. The Migrating Research Data Collections[9] project at the University of Michigan will examine research data migration between data management and preservation platforms over the lifetimes of specific datasets, while the Saving Data Journalism[10] IMLS grant will build on an existing open source computational reproducibility tool to prototype how it could be applied to help preserve data journalism and complex, interactive news websites. The Beyond the Repository project,[11] which builds on research conducted in 2017 about distributed digital preservation systems and is described more completely in Section 4.3, is building a curation toolkit that will enable a more efficient selection and management of materials for distributed digital preservation. While steady progress continues in these areas, organizational information sharing around the collection of content and access levels still needs greater community attention.

**Directions for 2025**

---

[6] DLF/HBCU Library Alliance, "Announcement of Partnership," 2018, https://www.diglib.org/groups/clir-dlf-affiliates/dlf-hbcu-library-alliance/.

[7] HBCU Library Alliance and Digital Library Federation, "Common Mission, Common Ground," 2017, https://www.diglib.org/wp-content/uploads/sites/3/2016/09/2017preconferencereport.pdf.

[8] DLF, "HBCU Library Alliance and Digital Library Federation launch "Authenticity Project" Fellowship Program," October 9, 2018, https://www.diglib.org/announcing-the-authenticity-project/.

[9] Andrea K. Thomer, "Migrating Research Data Collections (RE-07-18-0118),", 2018, https://www.imls.gov/sites/default/files/grants/re-07-18-0118-18/proposals/re-07-18-0118-18-full-proposal.pdf.

[10] "Saving Data Journalism," accessed April 8, 2020, https://savingjournalism.reprozip.org/.

[11] Weinraub (Northwestern University), "Beyond the Repository Grant Announcement," 2018, https://www.imls.gov/sites/default/files/grants/lg-70-18-0168-18/proposals/lg-70-18-0168-18-full-proposal.pdf.

One of the continued efforts within the digital curation community is the drive to approach the complexity and scale of digital collections via sustained, collaborative action. Collecting born-digital materials differs from collecting analog materials, but selection of materials has always been based on the same principles, regardless of format: selection of content for preservation over time is focused on making collection decisions that align with the strengths and mission of an institution. While some of these collecting decisions can be anticipated and planned for, identifying and collecting content during a rapidly unfolding event with large amounts of algorithm-driven, personalized, web-based content brings new and important challenges to curation. Collecting under these circumstances may sometimes begin without a clear sense of the scope and end point. Decisions about who will do the collecting, its scope, and how to collect that content responsibly and at a level of quality and transparency needed to support future research often need to be made closer to the time of content creation, before resources change or disappear altogether. These decisions benefit from broader discussions with content creators, researchers, and practitioners and actions needed to support them extend beyond our individual repositories. Collaborative action around collecting and preserving digital content is critical, with careful approaches to the development of the infrastructure of scaled digital collection that centers preservation intent, collection context, and user consent.

**Key Recommendations**

- Support sustained collaborative efforts (with practitioners, content creators, researchers, and more) to build, curate, share, and manage collections in ethically responsible ways in support of current and future research and accessibility;
- Develop and share strategies for targeted selection, preservation, and access to materials in the face of huge amounts of data that continue to grow not only with established formats, but newer and more complex digital materials; and
- Develop policies, practices, and community actions that acknowledge and reflect the realities of institutional priorities and resources, and the complex, time-consuming labor involved in building and sustaining digital content collections.

### 1.1.2 Organizational Policies and Practices

The 2015 *National Agenda* identified three recommendations related to Organizational Policies and Practices: advocate for resources; enhance staffing and training for digital stewardship; and foster multi-institutional collaboration.

**Where are we in 2020?**

These action items remain relevant and are ongoing today, a fact which is exemplified by significant shifts in collaborative organizations in both 2018 and 2019. In particular, the communication and staffing changes of the Digital Public Library of America and the community's response,[12] the cessation of publication of *D-Lib Magazine*,[13] as well as the operational retirement of the Digital Preservation Network (DPN)[14] in February 2019, underscore how fragile advocacy and multi-institutional collaboration can be.

Despite the vulnerability of the institutions supporting collective action in our community, collaboration has become recognized as an essential part of digital preservation. One demonstration of this is the generous support given by the Digital Library Federation (DLF) to the NDSA itself as it transitioned from its original home at the Library of Congress to DLF as a new host organization.

One positive trend is the creation of the Digital Preservation Services Collaborative,[15] an ongoing partnership among many non-profit digital preservation services including the Academic Preservation Trust (APTrust),[16] Chronopolis,[17] CNI, DuraSpace, Educopia, and more, who in 2018 released a declaration of shared values to underscore the need for collaborative stewardship and outline the professional ethics at the core of digital preservation work. This group remains active and continues to focus on future collaborative approaches to digital preservation.

**Directions for 2025**

Even though fears of dramatic budget cuts to federal granting agencies have not been fully realized, it is clear that reliance on external grant funding for ongoing stewardship activities is not sustainable. Further, many organizations are facing the heightened challenge of balancing priorities and allocating resources. Stewarding organizations need to be able to offer value in exchange for the resources required to successfully provide long-term digital stewardship.

---

[12] DPLA Board of Directors, "DPLA Board Response to Community Concerns," November 21, 2018, http://dpla.wpengine.com/wp-content/uploads/2018/11/DPLA_Board_of_Directors_Community_Letter_Response.pdf.

[13] Laurence Lannom, "The End of An Era," *D-Lib Magazine* 23, no. 7-8 (July/August 2017), http://www.dlib.org/dlib/july17/07editorial.html.

[14] "Information Update," The Digital Preservation Network, December 5, 2018, https://web.archive.org/web/20190226161550/http://dpn.org/news/2018-12-05-information-update.

[15] Digital Preservation Services Collaborative, "Declaration of Shared Values," accessed April 8, 2020, https://dpscollaborative.org.

[16] APTrust, accessed April 16, 2020, http://aptrust.org/.

[17] "Chronopolis," UC San Diego Libraries, accessed April 16, 2020, http://libraries.ucsd.edu/chronopolis/.

**Key Recommendations**

- Continue to collaborate and amplify work done by the preservation community (e.g. DPC's *Executive Guide*)[18] that provides tools and communication strategies for making the case for preservation.
- Create and share preservation advocacy templates and business plans with the broader community that represent multiple sectors.
- Explore and test models for sustainability of digital preservation training programs such as the National Digital Stewardship Residency (NDSR).[19]

### 1.1.3 Technical Infrastructure Development

In the area of Technical Infrastructure Development, the 2015 *National Agenda* identified two major recommendations: coordinate and sustain an ecosystem of shared services; and foster best practice development.

**Where are we in 2020?**

There is a strong synergy between technical infrastructure development and organizational coordination. Much of the technical infrastructure for digital preservation comprises open source tools, collaborative replication systems, and shared best practices and standards. For these to be successfully developed, deployed, and maintained, the coordination and contribution of resources by multiple organizations over substantial periods of time is required. The recent shifts in the landscape of digital stewardship organizations and services noted above could also negatively impact the long-term sustainability of the "ecosystem of shared services" recommended in 2015.

---

[18] Digital Preservation Coalition, "Executive Guide on Digital Preservation," accessed April 16, 2020., https://www.dpconline.org/our-work/dpeg-home.

[19] "About NDSR", National Digital Stewardship Residency Program, accessed April 8, 2020, https://ndsr-program.org/about/.

Despite these changes, several new efforts relevant to technical infrastructure have emerged in recent years. The OSSArcFlow project[20] is investigating and modeling a range of workflows for born-digital archival content, incorporating three leading open source software platforms: BitCurator, Archivematica, and ArchivesSpace. The Oxford Common File Layout (OCFL) specification,[21] which describes an application-independent approach to the storage of digital information in a structured, transparent, and predictable manner, is in an alpha release. It could be used in developing new digital repositories that incorporate promising new approaches for long-term object management. Also, emulation and virtualization are emerging as viable preservation and access strategies for born-digital materials. Grants from the Andrew W. Mellon and Sloan Foundations[22] are supporting the Scaling Emulation and Software Preservation Infrastructure (EaaSI) program[23] at Yale to expand the software preservation community and to develop open source software tools that enable the creation, management, and distribution of "virtual machines" which can simulate the hardware of an older computer on a newer computer and then run older software on the simulated machine.

**Directions for 2025**

In the realm of cultural heritage content, the rapid expansion of collection types such as moving image, web archives, and other large born-digital and digitized collections has revealed pain points for managing extremely large collections in a distributed service environment. Cloud computing has emerged as a crucial component in the ecosystem of shared services for digital preservation—but it is a component that creates its own preservation challenges.

Recent institutional collaborations represent an important step in the right direction. Strong partnerships and communication are required to successfully increase integration, interoperability, and collaboration among institutions, and across distributed application programming interfaces (APIs) and platforms.

---

[20] "OSSArc Flow," Educopia, 2017-2019, accesssed April 8, 2020, https://educopia.org/ossarcflow/.
[21] "Oxford Common File Layout," Oxford Common File Layout Community Group, accessed April 8, 2020, https://ocfl.io/.
[22] "Grants Database: Yale," Alfred P. Sloan Foundation, 2017, accessed April 8, 2020, https://sloan.org/grant-detail/8228.
[23] "About EaaSI," Software Preservation Network, accessed April 8, 2020, https://www.softwarepreservationnetwork.org/eaasi/.

**Key Recommendations**

- Continue to develop a matrix of services across the preservation landscape that encourages both transparency and awareness of the roles each service plays.
- Foster and refine the Levels of Preservation on a regular basis to keep pace with changes in digital preservation.
- Support emerging digital preservation formats and strategies via community efforts that aim to develop scalable, flexible technical and administrative infrastructures.
- Research, develop, and share digital preservation policies and workflows related to web applications, APIs, and cloud-based digital materials.
- Encourage digital repository systems and distributed digital preservation services to more broadly implement standards and best practices.

### 1.1.4 Research Priorities

The 2015 *National Agenda* identified two recommendations in the area of digital preservation Research Priorities: build the evidence base to justify the importance of digital preservation, and better integrate research and practice. Also, the 2015 report specifically identified several areas that required targeted research: cost modeling, environmentally sustainable preservation, computable significant properties, and trusted frameworks for stewardship.

**Where are we in 2020?**

Over the last five years, increased federal and private funding have supported medium to large efforts to research and develop digital preservation policies, planning, and practices. There have been a number of recent grants aimed at furthering the state of knowledge, implementation, education, and staffing of digital preservation efforts. A prime example is the systematic approach that the IMLS has taken to articulate and support a coordinated platform of research and practice.[24] While many of these projects do not have readily

---

[24] See the following publications: Trevor Owens, Ashley E. Sands, Emily Reynolds, James Neal, and Stephen Mayeaux, *The First Three Years of IMLS Investments to Enhance the National Digital Platform for Libraries*, (Washington DC: Institute of Museum and Library Services, Office of Library Services, 2017), https://digital.library.unt.edu/ark:/67531/metadc1259400/m2/1/high_res_d/imls-ndp-three-508.pdf and Trevor Owens, Ashley E. Sands, Emily Reynolds, James Neal, Stephen Mayeaux, and Maura Marx. "Digital Infrastructures that Embody Library Principles: The IMLS National Digital Platform as a Framework for Digital Library Tools and Services." (2018) *in Applying Library Values to Emerging Technology: Decision-Making in the Age of Open Access, Maker Spaces, and the Ever-Changing Library* (ACRL Publications in Librarianship #72),

measurable outputs as of yet, they have involved curators more deeply in the research process, and the efforts in this direction signal positive potential for digital preservation research and development.

Longitudinal surveys of community practice continue to provide an important part of the evidence base. NDSA reports[25] based on surveying the community, such as the *NDSA Web Archiving Report*, now published for the fourth time in 2018, the *NDSA Staffing for Effective Preservation,* published for the second time in 2017, and the *NDSA Storage Infrastructure Reports,*[26] now published for the third time in 2020, collectively comprise an important part of this evidence base. Looking across these survey results demonstrates the rapid growth in stewarded content, the incremental advances in practices and resources, and the substantial gaps in organizational preservation practice, resourcing, and planning.

**Directions for 2025**

Despite the promising developments summarized above, overall progress in key areas of preservation research is challenging to measure. There has been only incremental progress in closing the gap between the scale of information production and preservation, in validating trust models at all levels, and in developing testable models of future value and cost—including the environmental costs of long-term digital preservation.

Building a common evidence base around the disposition of stewarded content across the community has proven more difficult (in particular, for web-based materials) and relates to the challenges of specific workflow strategies as well as quality control for mass amounts of material. Although there has been an increase in the availability of preserved collections for domain research, these collections reveal the difficulty in making broad generalizations about stewardship practice across the community. It also highlights the knowledge gaps in the stewardship practice of many other collections.

Generally, case studies remain over-represented in digital preservation research, and articles in the field are less likely to contain highly-cited work than works in related computer- and information-science fields. Further progress will depend on the development of few rigorously validated preservation methods, wide-scale empirical

---

http://www.ala.org/acrl/sites/ala.org.acrl/files/content/publications/booksanddigitalresources/digital/9780838989401.pdf.

[25] Most NDSA reports are completed by NDSA Interest and Working groups and made available on the NDSA OSF site, https://osf.io/4d567/.

[26] Only the latest version of the Storage Infrastructure report is in the NDSA OSF repository, https://osf.io/uwsg7/. Review the appendix of the latest report for the locations of the previous two surveys and their data.

studies, probability-based surveys or field experiments, replicable simulation experiments, public test corpuses, testbeds, and recognized conformance tests.

**Key Recommendations**

- Funders should give priority to programs that systematically contribute to the overall cumulative evidence base for digital preservation practice.
- Funders should give priority to programs that are replicable and testable, and that rigorously integrate research and practice.

## 1.2 NDSA Role and Perspective

The overarching priorities identified in the previous section affect the entire preservation community and will require both independent and coordinated action by many stakeholders across the community. The NDSA strategic plan is strongly informed by the *Agenda*, and over the last five years, the NDSA and its partners have played an important part in building the evidence and practice base for the stewardship of digital content.

This contribution is exemplified by the longitudinal approach to examining digital preservation practice, as well as examining emerging topics to ensure that the baseline of community practice remains up to date and reflects ongoing changes to the field, technology, and the community. Since publication of the *2015 Agenda*, this work includes:

- The update of the NDSA Levels of Preservation via a Levels "reboot"[27]
- The formation of an NDSA Cloud Studies working group[28]
- The *NDSA Staffing for Effective Preservation Report,* published for the second time in 2017[29]
- The 2017 *NDSA Fixity Survey Report*, which grew out of the 2014 NDSA report *Checking Your Digital Content* and fixity recommendations in the 2014 and 2015 *NDSA National Agenda* reports[30]

---

[27] "History of the NDSA Levels of Preservation Working Group," NDSA, accessed April 16, 2020, https://ndsa.org/groups/levels-of-preservation/history/.

[28] "NDSA Cloud Studies working group," NDSA, accessed April 16, 2020, https://osf.io/zjyk8/.

[29] NDSA Staffing Survey Working Group. "Staffing for Effective Digital Preservation 2017," 2017, https://osf.io/mbcxt/.

[30] NDSA Fixity Survey Working Group. "2017 Fixity Survey Report," 2018, https://osf.io/grfpa/.

- The *NDSA Web Archiving Report,* published for the fourth time in 2018[31]
- Evolving preservation storage practices from the baseline of the 2011 and 2013 NDSA Storage Survey[32]
- Iterative efforts supported by both NDSA members and the broader digital preservation community—including the refined digital preservation storage criteria to support the work of digital preservation[33]

In addition to developing practices and building the community evidence base for preservation, NDSA member organizations participate in and rely on consortial preservation services and infrastructure that extend far beyond the NDSA itself. By design and necessity, these services and infrastructure extend across a large, international community. The broader stewardship community in both the United States and internationally has continued to build a strong base of consortial practice and capacity-building, including efforts around the stewardship of a wide variety of digital initiatives and collections—but this base requires continued commitment from the community.

# 2. Key Issues in Building Digital Content Collections

The 2015 *National Agenda* describes a gap in digital preservation between the types of materials created and used in our society and the types of materials that make their way into libraries and archives. Through progress in addressing this gap, examples of which are shared throughout this section, the community has gained practical experience and new perspectives on building, maintaining, and sharing digital content collections at scale and across a wide range of formats, as well as new understandings of areas where attention and priority are needed. Key issues include the need for sustained collaborative efforts (with practitioners, content creators, researchers, and more) to build, curate, and manage collections in ethically responsible ways to support current and future research and accessibility; the need to develop and share strategies for targeted selection in the face of overwhelming amounts of data in both established and new, complex formats;[34] increasing access and engagement with digital records as primary records;[35] and the very practical need to continue developing policies, practices, and community actions that acknowledge and reflect the realities of institutional priorities and resources and the complex, time-consuming labor involved in building digital content collections.

---

[31] NDSA Web Archiving Survey Working Group. "Web Archiving in the United States: A 2017 Survey," 2018, https://osf.io/ht6ay/.

[32] Michelle Gallinger et al., "Trends in Digital Preservation Capacity and Practice: Results from the 2nd Bi-annual National Digital Stewardship Alliance Storage Survey," *D-Lib Magazine* Vol. 23, No. 7-8, (July/August 2017), https://doi.org/10.1045/july2017-gallinger.

[33] The Preservation Storage Criteria was originally developed by Kate Zwaard, Gail Truman, Sibyl Schaefer, Jane Mandelbaum, Nancy McGovern, Steve Knight and Andrea Goethals in preparation for an iPRES 2016 workshop called "What is Preservation Storage?" Since then, Eld Zierau and Cynthia Wu have joined the original authors in working toward an improved version based on community feedback. Digital Preservation Storage Criteria Project, .accessed April 16, 2020, https://osf.io/sjc6u/.

## 2.1 Approaches to Content Selection at Scale

Collecting born-digital materials differs from collecting analog materials, but selection of materials has always been based on the same principle, regardless of format: selection of content for preservation over time is focused on making collection decisions that align with strengths and missions of an institution. Trevor Owens of the Library of Congress devotes an entire chapter in his book *The Theory and Craft of Digital Preservation*[36] to the intertwined relationship between preservation and collection development, and the fundamental need to clarify what content matters to an organization and to what the organization wishes to provide access in the future. The clarification of what matters varies across institutions, collections, and likely even specific content types. The process should include discussions on who should be making the decisions about what to save (should it be one "expert" or a collaborative endeavor? How should researchers and content creators be involved?), and how this impacts the long-term picture of whose voices are preserved. How an organization may choose to act or shift appraisal and selection policy in the face of environmental sustainability also plays an important role when considering questions of scale and selection.[37]

---

[34] Becker, Christopher, "Metaphors We Work By: Reframing Digital Objects, Significant Properties and the Design of Digital Preservation Systems," *Archivaria* 85 (2018: 6-36), https://tspace.library.utoronto.ca/bitstream/1807/87826/1/Metaphors%20We%20Work_TSpace.pdf. Owens, Trevor, "The Theory and Craft of Digital Preservation," LIS Scholarship Archive, July 15, 2017, doi:10.31229/osf.io/5cpjt.

[35] NARA, "2018-2022 Strategic Plan," February 2018, accessed April 8, 2020, https://www.archives.gov/about/plans-reports/strategic-plan/strategic-plan-2018-2022.

[36] Trevor Owens, *The Theory and Craft of Digital Preservation* (Baltimore: Johns Hopkins University Press, 2018), Chapter 5, "Preservation Intent and Collection Development" p. 81-102.

[37] Pendergrass et. al argue for a multi-faceted approach to environmental sustainability that includes paradigm shifts around appraisal, permanence, and availability of digital content. Considering environmental costs of digital preservation as part of appraisal is one of several proposed shifts in practice. Keith L. Pendergrass, Walker Sampson, Tim Walsh, and Laura Alagna (*2019*) "Toward Environmentally Sustainable Digital Preservation." *The American Archivist*, Vol. 82, No 1 (Spring/Summer 2019): pp. 165-206.

Examples of policy around collection development and selection of digital materials at scale include the Library of Congress's 2019-2023 Digital Strategy Document,[38] outlining how the "exponential growth" of digital collections will be met with an accelerated approach to digital content access while also improving search and protecting copyright owners, and strategic plans of the British Library (2015-2023)[39] and the National Library of Ireland (2016-2021),[40] which both expand on born-digital collection strategies. OCLC's 2017 *Research and Learning Agenda for Archives, Special, and Distinctive Collections in Research Libraries*[41] also highlights the "exponential" growth of born-digital materials, and distinguishes the need for structural work beyond the technical capture of digital content from carriers. For example, the report outlines a current need for a clear focus on the work *before* and *after* the technical transfer and ingest of digital content, including that of appraisal, selection and access at scale—in other words, being guided by and linking collecting efforts to development and stewardship policies.

For particular types of content, such as software or research data, it may be most useful to collaborate and build off of community and disciplinary methodology and priorities to aid in the application and development of broader organizational selection and appraisal approaches at scale. Workflow complexity and repetition is compounded when shared practice is difficult to leverage. Newly developed resources such as *Software Deposit Guidelines for Researchers*[42] and the 2019 Jisc research data study *What to Keep*,[43] which examines researcher practice and appraisal frameworks in the UK around the selection of research data for long-term preservation and access, provides two examples of subject expertise and content priorities that can be adapted into broader approaches to selection and collection development. Similarly, for archival and special collections, collaborating on and centering existing archival expertise that focuses on digital content context, appraisal, knowledge of legal issues, donor relations, authenticity, description, and content significance will also help build more scalable frameworks for content selection and collection development.[44] Recent open source projects such as Cobweb[45] aim to support selection at scale through collaborative collection development for thematic web archives, potentially allowing for distributed partners to make more informed local collection and capacity decisions around web archives and for the community to better identify gaps in collecting and stewardship. Researchers and practitioners such as computer scientist Alexander Nwala and others at Old Dominion University are also leveraging social media sources to support the selection of relevant web content to crawl at scale.[46]

---

[38] The Library of Congress, "Digital Strategy: the FY2019-2023 Digital Strategic Plan of the Library of Congress," April 26, 2019, https://www.loc.gov/static/portals/digital-strategy/documents/Library-of-Congress-Digital-Strategy-v1.1.2.pdf.

[39] The British Library, "Living Knowledge: The British Library 2015-2023," second edition, accessed April 8, 2020, https://www.bl.uk/britishlibrary/~/media/bl/global/projects/living-knowledge/documents/living-knowledge-the-british-library-2015-2023.pdf.

[40] The link to Strategic priorities for 2016-2021 is here: "Born Digital," National Library of Ireland, accessed April 8, 2020, https://www.nli.ie/en/born-digital-collection.aspx.

[41] Chela Scott Weber, "Research and Learning Agenda for Archives, Special, and Distinctive Collections in Research Libraries," ( Dublin, OH: OCLC Research. 2017), doi:10.25333/C3C34F.

[42] "Software Deposit Guidance for Researchers," The Software Sustainability Institute, August 7, 2018, accessed April 8, 2020, https://softwaresaved.github.io/software-deposit-guidance/.

[43] Neil Beagrie, "What to Keep: A Jisc research data study," February 2019, https://repository.jisc.ac.uk/7262/1/JR0100_WHAT_RESEARCH_DATA_TO_KEEP_FEB2019_v5_WEB.pdf.

The fundamental questions around collection development and selection policy are a particular challenge when selecting and collecting content during a rapidly unfolding event with large amounts of algorithm-driven, personalized web-based content. Collections may sometimes begin without a clear sense of the scope and duration of collection. Decisions about who will do the collecting, what should be collected, and how to collect that content responsibly and at a level of quality and transparency needed to support future research often need be made closer to the time of content creation, before ephemeral, web-based resources change or are no longer available.

While efforts like Documenting the Now currently lead the stewardship community in newer approaches to social media content collection, preservation, and a careful approach to concepts like user intent,[47] the infrastructure of scaled digital collection development that centers preservation intent, collection context, and user consent can also be found and further built on in other collections areas. Examples of this type of collections infrastructure range from permissions tools in use by the Library of Congress's web archiving program[48] to guidelines from George Washington Libraries around creating collection development policy for social media archives[49] to significant legal developments that allow for libraries, archives, and museums to circumvent technological protection measures on software in the pursuit of preserving software and the digital materials that depend on that software.[50]

---

[44] Jackie Dooley, "The Archival Advantage: Integrating Archival Expertise into the Management of Born-Digital Library Materials," OCLC Research, July 2015, https://www.oclc.org/content/dam/research/publications/2015/oclcresearch-archival-advantage-2015.pdf and Alex Chassanoff and Micah Altman,"Curation as "interoperability with the Future": Preserving Scholarly Research Software in Academic Libraries," May 23, 2019, https://doi.org/10.1002/asi.24244.

[45] "Cobweb," Regents of the University of California, accessed November 15, 2019, https://cobwebarchive.org/.

[46] Alexander Nwala, Michele C. Weigle, Michael L. Nelson, "Bootstrapping Web Archive Collections from Social Media," *HT '18*, July 9—12, 2018, Baltimore, MD, https://www.cs.odu.edu/~mln/pubs/ht-2018/hypertext-2018-nwala-bootstrapping.pdf.

[47] Greg Lyon and Mark Callahan, "Honoring User Intent on Twitter," August 17, 2014, accessed April 9, 2020, http://support.gnip.com/articles/honoring-user-intent-on-twitter.html.

[48] The Library of Congress, "Web Archiving Program Frequently Asked Questions for Site Owners," accessed November 15, 2019, https://www.loc.gov/programs/web-archiving/for-site-owners/frequently-asked-questions/.

### 2.1.1 Connection to Researchers

An ongoing and companion challenge of selection is how users interact with and use digital collections. Meeting the expanding needs of researchers with diverse skills and abilities needs to be addressed.[51] Researchers continue to seek better discovery and access as well as enhanced tools and options for the use and reuse of digital materials, including tools and workflows that ensure computational reproducibility.[52] Further, building connections to researchers provides institutions with opportunities to learn from their expertise and experience in specific content areas, identify gaps in collecting, and to consider new perspectives. Institutions can gain a better understanding of how materials are used, whether modes of access and description are useful (and accurate!), and where improvements in what and how we collect might be made. Research and new scholarship generated from our collections strengthens our understanding of our collections and the value and need to continue building and maintaining them.

Models for access have evolved as technology has advanced, and this advancement has been met with some renewed organizational priority, resourcing, and a greater focus on access to digital materials as core to missions of cultural stewardship, as well as a subject of study by digital stewards.[53] The Library of Congress Labs,[54] which launched as a way to encourage innovation with existing Library digital collections, provides researcher and public access to APIs, bulk content downloads, tutorials, and projects. The Collections as Data[55] project received IMLS funding to provide a strategic approach to developing, describing, providing access to, and encouraging the reuse of collections that support computationally-driven research and teaching. This work includes building functional requirements such as support for collections-as-data infrastructure development, developing use cases, and producing a collections-as-data framework that will ultimately encourage the computational use of digitized and born digital collections.[56]

---

[49] Social Feed Manager, "Building Social Media Archives: Collection Development Guidelines," updated March 13, 2017, accessed November 15, 2019, https://gwu-libraries.github.io/sfm-ui/resources/guidelines.

[50] Kendra Albert, "A Victory for Software Preservation: DMCA Exemption Granted for SPN," Harvard Law School Cyberlaw Clinic, October 26, 2018, https://clinic.cyber.harvard.edu/2018/10/26/a-victory-for-software-preservation-dmca-exemption-granted-for-spn/.

[51] See for example, New Media Consortium, "Digital Literacy in Higher Education, Part II: An NMC Horizon Project Strategic Brief," August 22, 2017, https://library.educause.edu/resources/2017/8/digital-literacy-in-higher-education-part-ii-an-nmc-horizon-project-strategic-brief and Society of American Archivists, "Guidelines for Accessible Archives for People with Disabilities," February 2019, https://www2.archivists.org/sites/all/files/SAA%20Guidelines%20for%20Accessible%20Archives%20for%20People%20with%20Disabilities_2019_0.pdf.

[52] Reprozip is an open source tool designed to aid with reproducibility of research by packaging all data files, environments, and libraries, https://www.reprozip.org/.

Experiments around the research use of web archives are instructive to stewards of all types of digital content. The Archives Unleashed project,[57] for example, is an effort to make petabytes of web archives data accessible to researchers for scholarly analysis, while also providing the tools and cloud computing environment necessary to do so. The project conducts a series of datathons: short and intensely focused meetings where participants lead discussions and get practical, hands-on experience with developing tools and data. This approach allows for outreach, education, and community building around researcher use of web archives.

Researchers stand to benefit as these and other new projects—including those that seek to describe more effectively and make born digital content discoverable,[58] encourage the access and reuse of web archives through research services,[59] or provide access at

---

[53] Rachel Appel, Alison Clemens, Wendy Hagenmaier, Jessica Meyerson, "Born Digital Access in Archival Repositories: Mapping the Current Landscape Preliminary Report," August 2015, https://docs.google.com/document/d/15v3Z6fFNydrXcGfGWXA4xzyWlivirfUXhHoqgVDBtUg/edit#heading=h.a6wak2j7q0pp.

[54] "Library of Congress Labs", accessed November 15, 2019, https://labs.loc.gov/.

[55] "Always Already Computational: Collections as Data," accessed November 15, 2019, https://collectionsasdata.github.io/.

[56] Thomas Padilla et al., "Always Already Computational: Collections as Data Final Report", Zenodo, May 22, 2019, doi:10.5281/zenodo.3152935.

[57] "About the Archives Unleashed Project," The Archives Unleashed Project, accessed November 15, 2019, https://archivesunleashed.org/about-project/.

[58] Example: "Guidelines for Born-Digital Archival Description version 1.0," University of California, October 26, 2017, https://github.com/uc-borndigital-ckg/uc-guidelines.

[59] Example: "Archive-It Research Services," Archive-It, accessed November 15, 2019, https://archive-it.org/blog/projects/archive-it-research-services/.

scale[60]—continue to develop around the practice related to and infrastructure necessary for the use of digital materials.

However, along with these developments, digital stewards must also be prepared to think about and structure the ways they can receive feedback from researchers about their collections, including researcher needs and expectations around the access and use of born-digital materials.[61] While efforts like the International Internet Preservation Consortium (IIPC) Research Working Group[62] provides an example of coordinated efforts to engage with the research community and share researcher use cases for web archives, it is notable that 49 percent of respondents to the 2017 NDSA Web Archiving survey indicated that they do not know if they have active researchers using their web archives at all.[63]

As recommended in the 2015 *National Agenda,* the work of stewards to examine and pursue lines of research and collaboration to better understand the use of stewarded collections and tools by researchers and users is ongoing. And while some progress in the form of research has been made in this area, more is needed.

### 2.1.2 Connection to Creator Community

Moving forward, digital stewardship organizations should continue to pursue activities within a wide variety of communities that produce, use, and conduct research with digital content in the course of their work or their daily life.

Industry-focused communities may have driving factors besides the needs of the users they serve. Business and financial incentives may not always translate to preservation best practices. However, there are instances where digital stewards play an important role in these communities. Examples of this active collaboration with communities of practice include the veraPDF project,[64] which was initially funded by the Preservation Formats for cultural information/e-archives (PREFORMA) and now led by the Open Preservation Foundation. An open source, industry supported PDF/A validator that was developed in partnership with the Validation Technical Working Group of the PDF Association as well as the Digital Preservation Coalition,[65] veraPDF represents a significant working partnership that is supported by the PDF software developer community and meets the needs for preservation by digital stewards.

---

[60] Example: "HTRC Architecture and Technical Organization," Hathitrust Digital Library, accessed November 15, 2019, https://www.hathitrust.org/htrc_architecture.

[61] Julia Kim, "Researcher Access to Born-Digital Collections: an Exploratory Study," *Journal of Contemporary Archival Studies*: Vol. 5 , Article 7 (2018), https://elischolar.library.yale.edu/cgi/viewcontent.cgi?article=1046&context=jcas and Tim Walsh, "CCA Access to Born-Digital Archives User Survey," May 4, 2017, https://www.bitarchivist.net/blog/2017-05-04-usersurvey/.

[62] "Research Working Group", International Internet Preservation Consortium, accessed April 9, 2020, http://netpreserve.org/about-us/working-groups/research-working-group/.

[63] Matthew Farrell, Edward McCain, Maria Praetzellis, Grace Thomas, and Paige Walker, "Web Archiving in the United States - A 2017 Survey," October 2018, https://osf.io/ht6ay/.

Additional examples include the Architecture, Design, and Engineering (ADE) summit[66] that took place at the Library of Congress at the end of 2017, where both content creators and caretakers (including architects, engineers, archivists, designers, government agencies, librarians, and many others) came together to discuss the state of the field, standards for new-build ADE assets, and case studies on current projects. Goals for the summit demonstrate the intertwined interests of both the ADE and digital stewardship communities, and express a desire for long-term sustainability, knowledge sharing, and promotion of best practices to allow for the interoperability of digital ADE assets. The continued growth in the use of 3D data and virtual or alternate reality calls for innovative tools and research methods within both the sciences and cultural history communities. Efforts to address some of those needs are reflected in initiatives such as Community standards for 3D data preservation (CS3DP)[67] and Developing Library Strategy for 3D and Virtual Reality Collection and Reuse.[68] Likewise, the The Andrew W. Mellon-funded Preserve this Podcast project[69] is an interesting approach to addressing digital stewardship within an intersection of communities that work in the same audio medium environment, yet represent different corporate, non-profit, and freelance production realities that have an impact on preservation and archiving practice.

---

[64] "VeraPDF," accessed on November 15, 2019, https://verapdf.org/home/.

[65] DPC, "Digital Preservation Coalition," accessed on April 29, 2020, https://www.dpconline.org/.

[66] "Designing the Future Landscape: Digital Architecture, Design & Engineering Assets," an event hosted by the Library of Congress, the National Gallery of Art and the Architect of the Capitol, November 16, 2017, http://digitalpreservation.gov/meetings/ade/ade2017.html.

[67] Jennifer Moore, "Community Standards for 3D data preservation (CS3DP)," IMLS grant, accessed November 15, 2019, https://www.imls.gov/sites/default/files/grants/lg-88-17-0171-17/proposals/lg-88-17-0171-17-preliminary-proposal.pdf.

[68] Nathan Hall, Jennifer Laherty, and Matthew Cook, "Developing Library Strategy for 3D and Virtual Reality Collection Development and Reuse," IMLS grant, accessed November 15, 2019, https://www.imls.gov/sites/default/files/grants/lg-73-17-0141-17/proposals/lg-73-17-0141-17-full-proposal-documents.pdf.

[69] Preserve this Podcast and Jacob Kramer-Duffield, "Podcast Preservation Survey Findings," February 5, 2019, http://preservethispodcast.org/assets/PodcastPreservation_SurveyFindings_Feb2019.pdf.

There are also many opportunities for the digital stewardship community to engage with communities outside of industry to preserve cultural heritage. For example, a cohort of twenty-seven public libraries, with the support of IMLS and the Internet Archive, is participating in the Community Webs program[70] to document local community histories. The creator community includes activist groups, individuals documenting their own family histories, professionals who wish to save their communications with colleagues, and hobbyists who build digital collections of objects reflecting their own interests. Some of the content created within these communities might be of value for preservation in cultural heritage institutions, and some communities and individuals may wish to manage their digital materials on their own with some level of support and guidance from the digital stewardship community.[71]

## 2.2 Content-Specific Challenges

In addition to the cross-cutting issues discussed above, specific forms of content pose urgent challenges to stewardship. Scientific data sets, dynamic web content, software, and massive collections of recorded video and audio pose specific technical and institutional questions that go beyond issues of the scale of content. This content is increasingly being recognized as a vital part of the scientific, cultural, and public record—but remains at high risk of loss.

### 2.2.1 Organizing and Ensuring Long-Term Access to Scientific Data Sets

Some of the most acute challenges of digital content can be illustrated by considering the curation of digital research data. The sheer scale of research data represents a daunting curatorial task. With newly developed scientific instruments and the growing use of computer simulations, a research team can generate many terabytes of data per day. Data curators face management at the petabyte scale and well beyond. Scientific fields such as particle physics, which generates collider data, and astronomy, which generates sky surveys, as well as research fields and methods like bioinformatics, crystallography, and engineering design produce massive amounts of digital data. By 2025, the total amount of genomics data alone is expected to equal or exceed totals from astronomy, YouTube, and Twitter.[72]

---

[70] "Community Webs," Archive-It, accessed November 15, 2019, https://archive-it.org/blog/projects/community-webs/.

[71] Bergis Jules, "Let the People Lead: Supporting Sustainability vs Dependency Models for Funding Community-Based Archives," *On Archivy,* November 3, 2017, https://medium.com/on-archivy/let-the-people-lead-supporting-sustainability-vs-dependency-models-for-funding-community-based-82f76d54c483.

Although some research data are no more complex than other objects that are routinely curated, a portion of digital research data are complicated to curate. Research data can be heterogeneous, ranging from numeric and image based, to textual, geospatial, and other forms. There are many different information standards used, as well as many different approaches to information structure (e.g., XML-structured documents vs. fixed image and textual file formats). Moreover, the research communities that produce data are equally diverse; data management practices vary greatly both within and between disciplines. Datasets are often not easily shared, findable, or interoperable, and scientists report spending much of their time data cleaning as opposed to creative tasks like mining data for patterns that lead to new research discoveries.[73] Predictable, interoperable data and data practices may also be of commercial interest.

Perhaps the overriding challenges common to all aspects of digital research data are the affiliated costs. This challenge was identified as a theme in a recent analysis of federal agency plans to ensure preservation and increase access to federally funded research data, and cited as an area that would benefit from greater detail and planning. Federal agencies plans showed that "without clear funding mechanisms, additional infrastructure development will be incremental at best."[74] Domain researchers, technologists, information scientists, and policymakers are searching for sustainable economic models with the ability to accurately predict costs and to balance them across the lifecycle (e.g., costs for ingest, archival management, and dissemination), and through federated inter-institutional repository systems.

One of the biggest needs for research data is the development of data management practices that reflect an understanding of the variations in data, whether it be raw, processed, summary, aggregate, preliminary, public use or metadata. Another research data need is a clear understanding of the variety of data uses, reuses, incentives, mandates

---

[72] Zachary D. Stephens, Skylar Y. Lee, Faraz Faghri, Roy H. Campbell, Chengxiang Zhai, Miles J. Efron, Ravishankar Iyer, Michael C. Schatz, Saurabh Sinha, and Gene E. Robinson, "Big Data: Astronomical or Genomical?" PLoS Biol 13(7): e1002195 (July 7, 2015), https://doi.org/10.1371/journal.pbio.1002195.

[73] CrowdFlower, "2016 Data Science Report," 2016, accessed November 18, 2019 https://visit.figure-eight.com/rs/416-ZBE-142/images/CrowdFlower_DataScienceReport_2016.pdf.

[74] Adam Kriesberg, Kerry Huller, Ricardo Punzalan, and Cynthia Parr. "An Analysis of Federal Policy on Public Access to Scientific Research Data," June 14, 2017, https://datascience.codata.org/articles/10.5334/dsj-2017-027/. [developed in response to a 2013 White House Office of Science and Technology Policy directive, the "Holdren memo"]

and responsibilities (including data security, integrity, and privacy protection), and the challenges of ebbs and flows of scientific funding. Of particular concern is the responsible management and sharing of data derived from human subjects, and the need to ensure adequate de-identification of participants.[75] The availability and accessibility of commercially produced data for social science research is another significant concern as the data on citizens, economy, and society is increasingly held by private businesses and only made available to researchers through special agreements.[76] There is no one-size-fits-all approach when it comes to resolving the management challenges of research data;[77] however, progress might be made by mobilizing the digital preservation and curation community toward in-depth study of these challenges of scale, complexity, research community practice, and cost with the aim of developing new recommendations and potential long-term solutions.

There have been notable areas of progress since publication of the 2015 *National Agenda*, including international meetings such as the First and Second Workshops on Scientific Archives bringing archivists, historians, scientists, and others together to discuss the appraisal, curation, and accessibility of science and technology archives.[78] The development and integration of the FAIR[79] guiding principles for research data stewardship into research policy and data management planning, including the National Institutes of Health Strategic Plan for Data Science[80] has been an important step in developing an infrastructure to support the reuse of data. There has been increased attention and 2020 NDSA Agenda for Digital Stewardshipcollaboration around ensuring long-term access to federal government scientific datasets led by groups such as the Environmental Data & Governance Initiative (EDGI), Data Refuge[81], and the End of Term (EOT) Web Archive,[82] as well as collaborations between researchers and private industry to support social science research using social media data, notably an April 2018 partnership between Social Science One and Facebook to explore "the effects of social media on democracy and elections."[83]

---

[75] National Institute of Health, "Request for Public Comments on a DRAFT NIH Policy for Data Management and Sharing and Supplemental DRAFT Guidance," November 8, 2019, https://www.federalregister.gov/documents/2019/11/08/2019-24529/request-for-public-comments-on-a-draft-nih-policy-for-data-management-and-sharing-and-supplemental.

[76] Henry Farrel, "There Aren't Any Rules on How Social Scientists Use Private Data. Here's Why We Need Them," *Items: Insights from the Social Sciences*, July 10, 2016, https://parameters.ssrc.org/2016/07/there-arent-any-rules-on-how-social-scientists-use-private-data-heres-why-we-need-them/.

[77] Liz Lyon, "Dealing with Data: Roles, Rights, Responsibilities and Relationships. Consultancy Report," Graham Prior, 2012 Managing Research Data, Facet Press, 2007, https://researchportal.bath.ac.uk/en/publications/dealing-with-data-roles-rights-responsibilities-and-relationships.

[78] Meetings emphasize both humanities and scientific reuse of contemporary scientific material, and importantly support a broader conception of the "scientific record," including laboratory notebooks, scientific diaries, correspondence, administrative records, etc, to document the history of science in addition to scientific research. 2018 Workshop: "Second Workshop on Scientific Archives," August 13-14, 2018, https://www.aip.org/second-workshop-scientific-archives.

[79] Mark D. Wilkinson, Michel Dumontier, IJsbrand Jan Aalbersberg, *et al.* "The FAIR Guiding Principles for scientific data management and stewardship," *Sci Data* 3, 160018 (2016), doi:10.1038/sdata.2016.18.

[80] The National Institutes of Health, "NIH Strategic Plan for Data Science," June 2018, https://commonfund.nih.gov/sites/default/files/NIH_Strategic_Plan_for_Data_Science_Final_508.pdf.

**Actionable Recommendations**

- Develop and refine tools to support at-scale curation using automated processes.
- Develop, share, and support management policies and practices that aim to ensure that data is FAIR (Findable, Accessible, Interoperable, and Reusable).
- Emphasize the need for value assessment to ensure that investments (of dollars, social capital, time, etc.) are informed by the evidence base (infrastructure, policies, grantee requirements, etc.) to ensure that resources are spent as efficiently as possible.

## 2.2.2 The Dynamic and Heterogeneous Data of the Web and Social Media

As noted in both the 2014 and 2015 *National Agendas,* web and social media continue to be areas of concern for preservation due to the exponential growth in the complexity and quantity of what constitutes the modern web. This area of focus is again included in this *NDSA Agenda* with both updated technical web archiving approaches and a deeper examination of the network of issues faced by the stewardship community tasked with preserving this type of content.

The mainstay tools of the web preservation community—the Heritrix archival crawler and the Wayback replay platform—are maladapted to the contemporary Web, because they are based on an increasingly outdated model of web content centered on static content and markup-based presentation.

However, the community has recently worked to develop crawling technology to address these issues. Brozzler,[84] a new crawler developed by the Internet Archive, aims to improve the capture of dynamic web content and is a promising new tool for users of the popular subscription-based Archive-It web archiving service.[85] Webrecorder,[86] a free and open source tool developed by Rhizome and funded by the Andrew W. Mellon Foundation, has also seen exponential growth in the web preservation community.[87] Much of the development in this area has focused on programmatic approaches to address the technical requirements for the capture of dynamic web content characterized by complex software interactions within browsers from Javascript execution to embedded video.[88] Tools like Webrecorder are able to archive user interactions with web resources in an effort to more faithfully capture dynamic web content (each page visited is included in the capture), but currently can be a challenge to implement at scale.

---

[81] "Data Refuge," accessed November 15, 2019, https://www.datarefuge.org/.
[82] End of Term Web Archive, "Project Background," accessed November 18, 2019, http://eotarchive.cdlib.org/background.html.
[83] "Social Science One," accessed November 18, 2019, https://socialscience.one/.
[84] "What is Brozzler," Archive-It Help Center, accessed November 18, 2019, https://support.archive-it.org/hc/en-us/articles/360000343186-What-is-Brozzler-.

While at a smaller scale for researcher access, desktop application development for the playback of local web archives files, such as the Webrecorder Player,[89] allows for more localized researcher access and review of multiple web archive formats without the need for an internet connection, which can aid access needs of a wide range of institutions.

According to the most recent NDSA Web Archiving Survey, many institutions reported the use of a combination of both Archive-It and Webrecorder tools, among other tools and services, to address the challenges of archiving in an increasingly dynamic web environment.[90] The heterogeneity of available tools, while a potential challenge for maintenance and workflows, can be seen as a useful development for the stewardship community. Similarly, support for interoperability between suites of tools based on best practices and format standards (including features such as the ability to upload ISO standard WARC files)[91] is a good step toward scalable, sustainable archives. The use of standard formats enables greater interoperability for playback and indexing across these popular web archiving tools, no matter their creation source. Ultimately, these standards provide for flexibility and adaptability of preservation and stewardship workflows and

---

[85] NDSA Web Archiving Working Group, "Web Archiving in the United States: a 2017 Survey," December 12, 2018, page 20 (94% of survey participants were Archive-It partners), https://osf.io/ht6ay/.

[86] "Webrecorder," accessed November 18, 2019, https://webrecorder.io/.

[87] NDSA Web Archiving Working Group, "Web Archiving in the United States: a 2017 Survey," December 12, 2018, page 21 (51 percent affirmed their use of Webrecorder), https://osf.io/ht6ay/.

[88] Programmatic approaches to web archiving also reflect the increasingly important role of browser emulation, especially for legacy technology, such as Adobe Flash https://rhizome.org/editorial/2018/dec/18/national-film-board-of-canada-and-rhizome/.

[89] "Webrecorder Player," WebrecorderGitHub, accessed April 9, 2020, https://github.com/webrecorder/webrecorder-player.

[90] NDSA Web Archiving Working Group, "2017 NDSA Web Archiving Survey Report," December 12, 2018, page 21, https://osf.io/ht6ay/.

[91] International Organization for Standardization, "ISO 28500:2017, Information and documentation — WARC file format," August 2017, https://www.iso.org/standard/68004.html.

reduce the risks that archived content will become uninterpretable as specific tools are replaced.[92]

The resource- and labor-intensive nature of web preservation work is an ongoing challenge for web archiving specialists and content stewards. Available web crawling tools continue to efficiently lower the barrier for technical capture, while the work of collection development, quality assurance, description, researcher access, and preservation remains difficult, given the resourcing realities of most United States cultural institutions, where often less than one full time employee is tasked with web archiving responsibilities.[93] Though structural guidance has recently been developed that will help address some of these existing resource barriers, particularly around descriptive practice,[94] continuing education,[95] and training,[96] these challenges remain.

These resourcing realities must be viewed in parallel with the growing interest of academics, web archive end-users, citizens, and other researchers to explore, extrapolate, and use web archives for everything from legal opinions[97] to large scale computational analysis. Though some progress has been made, there is a continued need for ongoing collaboration to understand and cultivate use cases for a wide variety of access and use needs of researchers for web archives. Forty-nine percent of respondents to the 2017 NDSA Web Archiving Survey indicated that they do not know if they have active researchers using their web archives.[98] The stewardship community needs to continue to build skills through collaboration with researchers and creator communities, (see this topic further covered in Sections 2.1.1 and 2.1.2), that will enable the testing, development and refinement of new analytical tools and methods.

---

[92] Both Webrecorder and Archive-It now support the ability to upload WARC files from any source, see "Integrate external WARC/ARC files into Archive-It collections," Archive-It, accessed November 18, 2019, https://support.archive-it.org/hc/en-us/articles/360000651246-Integrate-external-W-ARC-files-into-Archive-It-collections.

[93] NDSA Web Archiving Working Group, "2017 NDSA Web Archiving Survey Report," 13-15, https://osf.io/ht6ay/.

[94] Jackie Dooley and Kate Bowers, "Descriptive Metadata for Web Archiving: Recommendations of the OCLC Research Library Partnership Web Archiving Metadata Working Group," Dublin, OH: OCLC Research, (2018), https://doi.org/10.25333/C3005C.

[95] "Continuing Education to Advance Web Archiving (CEDWARC) project," accessed November 25, 2019, https://cedwarc.github.io/. The CEDWARC project develops a continuing education curriculum and teaches library and archive professionals advanced web archiving and analysis techniques.

[96] "Training Working Group," International Internet Preservation Consortium, accessed November 25, 2019, http://netpreserve.org/about-us/working-groups/training-working-group/.

[97] "Perma.cc," accessed November 18, 2019, https://perma.cc/.

[98] NDSA Web Archiving Working Group, "2017 NDSA Web Archiving Survey Report," 23, https://osf.io/ht6ay/.

As web preservation efforts continue to develop in regard to technical collection and researcher use, some important related research and discussions around the concepts of authenticity and trust,[99] appraisal and selection,[100] and provenance[101] for web archives have deepened. There is also ongoing research into the practice and importance of preserving and evaluating web archives as public records, particularly in the field of journalism.[102]

Web archives work has continued to develop in ways that reflect the equally important roles of communities, activism, and societal issues in addition to technical developments. In March 2018, Rhizome, in collaboration with the DocNow team, led a National Forum on Ethics and Archiving the Web[103] where web archivists, technologists, artists, librarians, and activists shared experiences and reflections on wide ranging topics such as documenting traumatic events, like the 2017 "Unite the Right" rally and counter-protests in Charlottesville, Virginia; web archiving as a civic duty; and broadly, the politics and power behind decisions about what and whose histories are preserved on these platforms. These types of conversations are opportunities to engage with peers who continue to be valuable advocates around web preservation and data ethics practice, including social scientists, legal scholars, computer and data scientists, activists, journalists, and others who hold significant experience framing and dealing with ethical,[104] policy, and technical issues around social media data, and the digital stewardship community should continue to work on meaningful partnerships with these communities.

Social media content in particular has become increasingly important for understanding policy, for legal research, and for social science, and is particularly powerful in terms of allowing for computational analysis on machine-actionable, structured data.[105] Social media epitomizes the mismatch between the contemporary Web and the web crawling paradigm, exacerbates concerns over privacy, ethics, and copyright, and highlights not-easily-resolvable tensions between corporate business models and cultural heritage organizations' interest in preserving and providing access to data that reflects cultural history. While there has been some movement in the stewardship and technical communities to more closely align API-based social media collecting with web archiving to better aid future research, more development is needed.[106]

---

[99] Mohamed Aturban, Michael L. Nelson, and Michele C. Weigle, "Difficulties of Timestamping Archived Web Pages," December 8, 2017, https://arxiv.org/pdf/1712.03140.pdf.

[100] Edward Summers, "Appraisal Practices in Web Archives." *SocArXiv*, March 15, 2019, doi:10.31235/osf.io/75mjp.

[101] Emily Maeumura, Nicholas Worby, Ian Milligan, and Christoph Becker, "If these Crawls Could Talk: Studying and Documenting Web Archives Provenance," *Journal of the Association for Information Science and Technology*, Vol. 69, no 10 (October 2018), https://doi.org/10.1002/asi.24048.

[102] Sharon Ringel and Angela Woodall, "A Public Record at Risk: The Dire State of News Archiving in the Digital Age," Tow Center for Digital Journalism at Columbia's Graduate School of Journalism, March 28, 2019, https://www.cjr.org/tow_center_reports/the-dire-state-of-news-archiving-in-the-digital-age.php.

[103] "The National Forum on Ethics and Archiving the Web," March 22-24, 2018, https://eaw.rhizome.org/.

[104] Matthew L. Williams, Pete Burnap, and Luke Sloan, "Towards an Ethical Framework for Publishing Twitter Data in Social Research: Taking into Account Users' Views, Online Context and Algorithmic Estimation," Sociology, 51(6), (2017): 1149—1168, https://doi.org/10.1177/0038038517708140.

Access APIs for social media, even those that are currently publicly available, represent a degree of commercial service provider-sanctioned control over digital content that is of great concern to digital stewards who aim to preserve and provide access to these types of collection materials over time. Public APIs help navigate some technical collecting challenges, but also reflect a troubling shift from the adequacy of a general-purpose tool to the prospect of having to devise individualized strategies for each commercial platform, along with competing business models from those same platforms that increasingly rely on selling access to data. Moreover, the economic concentration of social media platforms creates the threat of sudden loss of access, whether for economic reasons or political ones.[107] Over the last four years, API access to major social media platforms has become increasingly restricted, and access to Twitter and Facebook APIs has been drastically curtailed in 2018 in the wake of political scandal[108]—with no carve-outs for preservation. Further, whether obtained through an API or through scraping, the legal risks of providing long-term access to social media data are evolving and complex.[109]

Social media content is also more concentrated than web content in that it is primarily controlled by a handful of large companies. One implication of this concentration is that it is more difficult for small and medium-sized archives to coordinate on systematically archiving social media because of the large degree of local resources and capacity-building that would be needed. Even larger governmental or academic organizations that have the capacity to collect and archive the data may not have the capacity to offer access and use of that data to the community—a lesson illustrated by the Library of Congress's foray into Twitter archiving, now discontinued.[110]

---

[105] Sara Day Thompson, "Preserving Social Media," DPC Technology Watch Report 16-01, February 2016, DOI: http://dx.doi.org/10.7207/twr16-01.

[106] Justin Littman, Daniel Chudnov, Daniel Kerchner, Christie Peterson, Yecheng Tan, Rachel Trent, Rajat Vij, and Laura Wrubel, "API-Based Social Media Collecting As a Form of Web Archiving," International Journal on Digital Libraries (2018) 19: 21, https://doi.org/10.1007/s00799-016-0201-7.

[107] A recent example would be Verizon's discontinuation of Yahoo Groups and even thwarting attempts to preserve that content: "Verizon kills email accounts of archivists trying to save Yahoo Groups history,"December 9, 2019, accesssed April 9, 2020, https://www.zdnet.com/article/verizon-kills-email-accounts-of-archivists-trying-to-save-yahoo-groups-history/.

[108] Bob Reselman, "It's the End of the API Economy As We Know It," July 5, 2018, accessed April 9, 2020, https://www.programmableweb.com/news/its-end-api-economy-we-know-it/analysis/2018/07/05.

[109] David O'Brien, Jonathan Ullman, Micah Altman, Urs Gasser, Michael Bar-Sinai, Kobbi Nissim, Salil Vadhan, Michael Wojcik, and Alexandra Wood, "Integrating Approaches to Privacy Across the Research Lifecycle: When Is Information Purely Public?" Berkman Center Research Publication No. 2015-7,March 27, 2015, http://dx.doi.org/10.2139/ssrn.2586158.

While the investments and progress in technical web harvesting and collecting tools have been substantial over the last several years, further work is needed to match this progress in terms of community development around sustainable workflows and ethics and privacy frameworks for web and social media collection, as well as scalable quality assurance, researcher and community use, and enhanced collection development and appraisal, description, and access workflows.

**Actionable Recommendations**

- Increase and advocate for sustained institutional and organizational support for the ongoing and significant labor of preserving web and social media, including web archives selection, quality control at scale, description, policy development, preservation, and access.
- Increase collaboration with researchers and communities who use or seek to collect web and social media archives to better understand their needs and cultivate use cases, as well as to encourage researcher and community use of new access models and methods, including consideration of ethical and privacy frameworks and intersections with commercial platforms.
- Continue to support the development and maintenance of open source web archiving tools that can effectively capture dynamic web content and social media.
- Increase collaboration around communities active and involved with web and social media archives and ethical issues, including journalists, activists, social scientists, legal scholars, artists, and technologists.
- Engage with communities of web producers, service providers and creators, especially social media companies, to instill the value of archivability in their terms of use and design.

### 2.2.3 Scaling Software Preservation

It is essential that the digital stewardship community continues to make strides toward the preservation of software. Software, for the purposes of this *Agenda*, is meant to refer both to applications and system software specifically (including operating systems, device drivers, etc.) as well as the networks and workflows that also contain its context, components, and meaning. Software is now fundamental to science, as the majority of scientific claims are now critically dependent on applying software to evidence. Much of what is considered web content, particularly with HTML5, is essentially software, and the authentic preservation of web content will rely on approaches to software preservation.[111] Software touches nearly all components of modern culture, from artwork to engineering. As both a key to providing continuing access to the digital objects that depend on software, and as an important cultural artifact in its own right, software represents a significant challenge and stewardship opportunity for the community.

---

[110] Gayle Osterberg, "Update on the Twitter Archive at the Library of Congress," Library of Congress Blog December 26, 2017, https://blogs.loc.gov/loc/2017/12/update-on-the-twitter-archive-at-the-library-of-congress-2/.

Software preservation has been a significant area of collaboration within international community efforts. These community efforts are necessary to approach the complex legal and infrastructural environments for software preservation. In the United States, several recent developments are crucial to the baseline establishment of the necessary legal and organizational environments required for the large-scale challenge of the preservation of software and software-dependent materials.

This progress has been marked by community research and action such as the Association for Research Libraries' report, *The Copyright Permissions Culture in Software Preservation and Its Implications for the Cultural Record*[112] which provides a much needed legal examination of software preservation from an informed fair use perspective. A second report, *Code of Best Practices in Fair Use for Software Preservation,*[113] was issued in the fall of 2018, and builds on these initial findings and research to form guidelines from which preservation professionals will be able to make informed decisions for their preservation work.

In concert with these guiding reports, 2018 also saw successful legal action in the United States in the form of a Digital Millennium Copyright Act (DMCA) exemption for software preservation.[114] This legal reality has since been distilled into *A Preservationist's Guide to DMCA Exemption for Software Preservation*,[115] which aims to guide preservationists, librarians, and others in the cultural heritage community to make educated decisions about how their activities may fall under this exemption.

---

[111] See the "HTML 5.2 WC3 Recommendation," December 14, 2017, https://www.w3.org/TR/html52/ and technical detail in Mohamed Aturban, Michael L.Nelson, and M. C. Weigle, "Difficulties of Timestamping Archived Web Pages," December 8, 2017, https://arxiv.org/pdf/1712.03140.pdf both accessed April 9, 2020.

[112] Patricia Aufderheide, Brandon Butler, Krista Cox, and Peter Jaszi, "The Copyright Permissions Culture in Software Preservation and Its Implications for the Cultural Record," Association of Research Libraries, February 9, 2018, https://www.arl.org/resources/the-copyright-permissions-culture-in-software-preservation-and-its-implications-for-the-cultural-record/.

[113] Patricia Aufderheide, Brandon Butler, Krista Cox, and Peter Jaszi, "Code of Best Practices in Fair User for Software Preservation," September 2018, revised February 2019, https://www.arl.org/resources/code-of-best-practices-in-fair-use-for-software-preservation/.

These efforts in the United States join the groundbreaking international progress of the Software Sustainability Institute (SSI)[116] around the preservation, advocacy, and sustainability of research software, and the efforts of the Software Heritage archive (SH)[117] to collect, preserve, and share publicly all available source code and development documentation. These large scale, multifaceted efforts have also recently been supported by the articulation of use cases and infrastructures, such as the documentation and adoption of practices by research communities to evaluate and preserve software related to computer science research, and the specific software preservation use cases provided by the cultural heritage sector for the long comment related to the DMCA exemption. These combined efforts will allow the cultural heritage field in the United States to move from the stagnation of legal and administrative uncertainty toward informed, active preservation practice of software and software-dependent works.[118]

Scalable technical infrastructure for use and access is a necessary companion to understandable legal environments for the work of software preservation. Progress in emulation frameworks and community development have moved forward in ways that are beginning to outline a technical approach for standardized pathways to the preservation and access of software and software dependent materials.[119] Grant-funded projects at Yale[120] and the Software Preservation Network (SPN)[121] aim to address the specific technical, legal, metadata, and access challenges for software and software-dependent materials preservation, while also supporting the work of building a technical infrastructure and engaging the broader digital preservation community around the need for software preservation. To this end, SPN, Cal Poly and IMLS are also supporting a small cohort of libraries, museums, and archives to examine institutional software preservation use cases and to both broaden participation in software preservation and advance preservation practice.[122] Scalable technical infrastructure also touches on discoverability and access. In addition to the work of the SH to collect and preserve all publicly available source code, including code in GitHub,[123] GNU,[124] GitLab,[125] HAL[126], Google Code[127] and Inria[128], the organization also recently had a public release, and has introduced search functionality via web browser or web API to allow anyone to check to see if their code has been archived.[129]

---

[114] "Software Preservation Comments Filed in 1201 Rulemaking," Harvard Law School Cyberlaw Clinic, January 2, 2019, https://clinic.cyber.harvard.edu/2018/01/02/software-preservation-comments-filed-in-1201-rulemaking/ and Library of Congress U.S. Copyright Office, "37 CFR Part 201: Exemption to Prohibition on Circumvention of Copyright Protection Systems for Access Control Technologies," Federal Register, Vol. 83, No. 208, October 26, 2018, https://www.govinfo.gov/content/pkg/FR-2018-10-26/pdf/2018-23241.pdf.

[115] Kee Young Lee and Kendra Albert, "A Preservationist's Guide to DMCA Exemption for Software Preservation," December 10, 2018, http://www.softwarepreservationnetwork.org/1201-exemption-guide-for-software-preservationists/.

[116] "About the Software Sustainability Institute," Software Sustainability Institute, accessed April 9, 2020, https://www.software.ac.uk/about.

[117] "Opening the archive, one API (and one FOSDEM) at a time," Software Heritage, February 4, 2017, accesssed April 9, 2020, https://www.softwareheritage.org/archive/.

[118] See e.g. "Software and Data Artifacts in the ACM Digital Library," Association for Computing Machinery, accessed August 30, 2019, https://www.acm.org/publications/artifacts and "UNESCO call for recognition of software source code as an important component of heritage for sustainable development" UNESCO, November 16, 2018, accessed April 9, 2020, https://en.unesco.org/news/experts-call-greater-recognition-software-source-code-heritage-sustainable-development.

Because of these large-scale legal and technical advances in software preservation over the past years and the tireless efforts of many in the community, great strides have been made in software preservation over the last few years. Yet, a good deal of more focused implementation work remains to be done to build on these efforts. The NDSA would like to see a range of institutions take or build on the following actions to continue the community progress for software preservation.

---

[119] The Emulation as a Service project ( http://eaas.uni-freiburg.de/) and Emulation as a Service Infrastructure  https://www.softwarepreservationnetwork.org/eaasi/) are implementing scalable emulation service and technical models.
[120] Cummings, Mike, "Project Revives Old Software, Preserves 'Born-Digital' Data," February 13, 2018, https://news.yale.edu/2018/02/13/project-revives-old-software-preserves-born-digital-data.
[121] Software Preservation Network, "Affiliated Projects," accessed on April 29, 2020, http://www.softwarepreservationnetwork.org/projects/.
[122] "FCoP Overview," Fostering a Community of Practice Software Preservation Network, accessed April 9, 2020, http://www.softwarepreservationnetwork.org/fcop/.
[123] "GitHub," accessed April 17, 2020, https://github.com.
[124] "GNU Operating System," accessed April 17, 2020, https://www.gnu.org/home.en.html.
[125] "GitLab," accessed April 17, 2020, https://about.gitlab.com/.
[126] "Hyper Articles en Ligne (HAL)" accessed April 21, 2020, https://hal.archives-ouvertes.fr/.
[127] "Google Code," accessed April 17, 2020, https://code.google.com/.
[128] "Inria" accessed April 21, 2020, https://www.inria.fr/en
[129] "Search Archived Software," Software Heritage Archive, accessed April 9, 2020, https://archive.softwareheritage.org/browse/search/.

**Actionable Recommendations**

- Outreach and engagement with the software industry, organizations such as the National Institute of Standards and Technology and National Software Registry Library, professional societies such as the Association for Computing Machinery, the American Institute of Architects, and the Institute of Electrical and Electronics Engineers, as well as the hobbyist and gaming communities should continue to better share efforts to build capacity around issues of software preservation.
- Consortial efforts that aim to address the large scale, multifaceted problem of software preservation such as the SPN should continue to be supported and sustained, and should coordinate with related international efforts such as SSI and SH when possible.
- Legal environments and guidelines for the collection of, preservation of, and access to software need to be tested, shared, and implemented in a variety of institutional contexts.
- Continued investments in scaling research and tool development for virtualization and emulation of computing environments, like the bwFLA or WebAssembly,[130] are necessary to make software and software-dependent objects usable over time.
- Research regarding technical standards for the creation of standardized formats for software images, and metadata that allows for emulation and other methods of scalable access for reproducibility to work needs to be developed and tested to allow for quality assurance, interoperability and machine-actionable processing.[131]
- Institutional ingest, collection, creation and preservation environments for various software types and software-dependent objects across cultural heritage institutions—from donor agreements to data management plans to archival processing and user access—need to be researched, documented, and tested.

### 2.2.4 Scale and Complexity of Moving Image and Recorded Sound Data

Digital preservation and stewardship of motion picture film, audio, and video presents a multitude of challenges. There is a need for both new standards and for the evolution of existing standards, such as preservation-quality reformatting and a myriad of issues that arise from creating and managing large files—not only storage, but the long-term ability to manage and play back these files. While the motion picture and recording industries *should* collaborate with cultural heritage institutions, this is not always the case. It is vital that both content creators and stewards work together to develop standards and workflows that will ensure long-term access to our recorded and moving image heritage. The cultural heritage community must continue to engage and encourage private/commercial and institutional relationships. In many cases, the tools and applications needed for both environments are useful for each community.

---

[130] "bwFLA-Emulation as a Service," bwFLA, http://eaas.uni-freiburg.de/ and "WebAssembly," WebAssembly, https://webassembly.org/, both accessed April 9, 2020.

[131] Wikidata is an example of a technical registry that attempts to address this type of need using linked open data concepts. Katherine Thornton, Euan Cochrane, Thomas Ledoux, Bertrand Caron, and Carl Wilson, "Modeling the Domain of Digital Preservation in Wikidata." In Proceedings of ACM International Conference on Digital Preservation, Kyoto, Japan, September 2017 (iPres'17), 10 pages, https://ipres2017.jp/wp-content/uploads/7.pdf.

The ease of digital media creation has erupted due to the proliferation of easy-to-use cameras, each creating a wide variety of file format outputs. The digital preservation systems and infrastructure must be able to accommodate the ever-growing list of file formats to allow efficient preservation, access, and migration, whether that be transcoding to a single format or the ability to store and retrieve many native formats. Final products are very large media files, and increasing in resolution and size. Large files take more time to move, copy, process, and store, which means more resources for computing power, storage, people, and time. The "more" theme is very present. The migration process alone becomes overwhelming and all consuming.

Interest in cutting edge research to utilize computational tools to help with cataloging, discovery, and organization of digital media material is growing. As more media content can be analyzed as data, digital humanities researchers are utilizing data sets of imagery and media for research. It is bringing together the computer science experts with cultural institutions and the humanities—disciplines that are now learning from each other. Funding for this collaborative research is increasing, but has yet to be matched by funding for the preservation of these materials.

Finally, analog media created over the last fifty to sixty years is deteriorating at a rapid rate. Video tape and sound recording formats are becoming obsolete—the equipment needed to play the formats is disappearing, required parts are no longer manufactured, and the physical tape itself is deteriorating. Digitizing analog materials is now considered the best preservation strategy and the best method for new distribution and access. This adds to the increasing collections of digital media files that require long-term preservation. As a preservation practice, this extensive conversion to digital scales poorly and must be addressed immediately.[132]

Preservation funding for audio visual materials is extremely limited. Given the volume at risk, and the short time frame in which to preserve the materials, there is not enough easily accessible, public funding available to save the volume of content created in the last sixty years for just public media. Much of the available funding for analog to digital migration goes towards paper and photographs. The number of items that can be digitized for those formats is far greater than audiovisual materials. Therefore the dollars go farther, and may be seen as a greater value. But those formats will last far longer than the magnetic tape that audiovisual content has been stored on. Paper and photographs, if stored properly, can survive over 100 years. Magnetic tape will last maybe sixty years if lucky, and we have reached the sixty-year age limit for the earlier video recordings. Newer, cheaper tape formats last for less time.[133]

---

[132] Mike Casey, "Why Media Preservation Can't Wait: The Gathering Storm," 2015, https://www.avpreserve.com/wp-content/uploads/2015/04/casey_iasa_journal_44_part3.pdf

In addition, creation of born-digital audiovisual content is increasing. There are solutions for storage and short-term management of these files during the production creation process, but standardized workflows and processes of what to do with these files for preservation is lacking. Much of the material produced gets stored on hard disks and placed on a shelf to be dealt with later. In addition, once a work is 'published' or broadcast, the final digital copy is often not captured - particularly if the work is being published on the web.

Rights issues are another complicating factor for preserving recorded sound and moving images. They are not covered currently under the exceptions in Section 108 of the Copyright Law, meaning all non-text-based works cannot explicitly be copied for preservation purposes, though they can be copied for fair use purposes.[134] For sound recordings alone, rights considerations need to be given to the recording itself, the performer, and the writer of the music. Films and television programs have distributors, copyright owners, talent unions, and owners of third-party materials used in a final program or film—music, stills, historic footage. Where is the responsibility to preserve this content and provide it to the necessary agencies/unions needed? Who can legally preserve it? Many funders will not help support preservation efforts unless public access is promised. How can an institution fulfill that promise when the rights issues for access are so complicated? Where are the gaps in this process that Congress should address, considering the divergent (and declining) revenue streams of content creators and content stewards and the fact that intellectual property owners have strong lobbies?

Much of the 20th and 21st centuries' cultural heritage and history is documented on audiovisual media. As a democratic nation that sees the importance of understanding the past as we look to the future, it is important to find solutions for the long-term preservation and storage of, and access to, these materials. As a result, there is a significant need to further define and communicate what preservation formats are in this area and what workflows should look like for working with and maintaining the authenticity of increasingly complex forms of digital audio and video files.

---

[133] Joshua Ranger, "The Cost of Inaction," AVP Blog, February 12, 2013, https://www.weareavp.com/the-cost-of-inaction/.

[134] Section 108 Study Group, "The Section 108 Study Group Report," Library of Congress, March 2008, http://section108.gov/docs/Sec108StudyGroupReport.pdf.

**Actionable Recommendations**

- Engage and encourage relationships between private and commercial and heritage organizations to work together to develop standards and workflows that will ensure long-term access to our recorded and moving image heritage.
- Support the ability of digital preservation systems and infrastructure to accommodate the ever-growing list of file formats to allow efficient preservation, access, and migration.
- Explore options for dealing with difficult intellectual property rights for preserving recorded and moving image materials.
- Advocate for funding to migrate analog formats to digital formats before the content is lost.

## 2.2.5 Computational Techniques for Managing Digital Materials

The potential loss of digital material that represents modern culture, scholarship, and government continues to pose a threat to the greater cultural memory. From email and scholarly datasets to government archives and medical research, organizations aim both to apply focused action to the preservation of large amounts of digital material, as well as to take the necessary approaches to then render these digital materials discoverable and accessible over time.[135] The National Institutes of Health (NIH), in the organization's first strategic plan for Data Science in 2018, framed the issue succinctly: "The growing costs of *managing* data could diminish NIH's ability to enable scientists to *generate* data for understanding biology and improving health."[136]

Yet the exponential growth of digital material creation and the associated needs for preservation and access have been met with some notable developments. In the archival field, this includes a broader strategy for research and action around concepts such as computational access to records, concerns about privacy, the application of forensic techniques, and development of collaborative infrastructures to deal with huge numbers of born digital records,[137] as well as a call for the formal articulation of a new transdiscipline of computational archival science (CAS). CAS is an interdisciplinary approach that strives to apply computational methods and resources to support large-scale processing, analysis, and preservation of, as well as access to, digital archives with the aim of improving efficiency, productivity and precision in support of appraisal, arrangement and description, preservation and access decisions, and engaging and undertaking research with archival material.[138] One potential implication of the CAS approach and modern records management as a whole is that rather than relying on file clerks to organize and store information, the information creator—each institution and individual—will be responsible for properly managing his or her own electronic records.[139] A proper infrastructure, supplemented with public outreach, will be critical in educating the public about current deficiencies in long-term electronic preservation and in equipping them to properly save their important materials. Personal digital archiving and community archives also continue to play a fundamental and growing role in outreach, collaboration and feedback around these topics.[140]

---

[135] See the National Archives and Records Administration Strategic Plan 2018-2022 for an example addressing the need for scalable processing and access to huge record volumes. NARA, "2018-2022 Strategic Plan," February 2018, accessed April 8, 2020, https://www.archives.gov/about/plans-reports/strategic-plan/strategic-plan-2018-2022.

[136] The National Institutes of Health, "NIH Strategic Plan for Data Science," June 2018, https://commonfund.nih.gov/sites/default/files/NIH_Strategic_Plan_for_Data_Science_Final_508.pdf.

[137] Wener, Chela Scott. (2017). Research and Learning Agenda for Archives, Special, and Distinctive Collections in Research Libraries. Dublin, OH: OCLC Research. doi:10.25333/C3C34F

Over the last several years, the stewardship community at large has also made progress in the application of computational techniques to help address the modern scale and access needs surrounding digital materials. The efforts include the use of natural language processing and named entity recognition techniques drawn from the fields of computer science and linguistics in use by projects such as ePADD,[141] an IMLS-funded project to address the issues around the appraisal, processing, preservation and access to email records. These computational techniques address processing realities related to the sheer volume of material, as well as approaches as to how to address privacy, duplication, and access concerns around modern correspondence at scale. Related projects using

---

[138] Marciano, Richard, Victoria Lemieux, and Mark Conrad. "Archival Records and Training in the Age of Big Data." In *Re-Envisioning the MLS: Perspectives on the Future of Library and Information Science Education*, Volume 44B:179–99. Advances in Librarianship. Emerald Publishing Limited, 2018. https://dcicblog.umd.edu/cas/wp-content/uploads/sites/13/2017/06/Marciano-et-al-Archival-Records-and-Training-in-the-Age-of-Big-Data-final.pdf.

[139] For an illustrative example, the 2011 NARA Records Management Self-Assessment report states, "to have an effective records management program, agency records management staff must have a baseline of knowledge about electronic records and how to manage them. Records staff do not need to be technological experts, but they have to understand certain fundamental principles and practices of managing electronic records." National Archives and Records Administration. "2011 Records Management Self-Assessment Report," 2011. https://www.archives.gov/files/records-mgmt/resources/self-assessment-2011.pdf

[140] For an example of the role of community archives and collaboration in their own archives and records management, see the 2017 report from the Cultural Heritage and Social Change Summit: "Nothing About Us Without Us," Cultural Heritage and Social Change Summit, accessed April 6, 2020, https://about.historypin.org/content/uploads/2017/12/HistoryPin_CHSC_takeaways_final.pdf.

[141] LG-70-15-0242-15, Institute of Museum and Library Services, accessed April 6, 2020, https://www.imls.gov/grants/awarded/lg-70-15-0242-15

computational and statistical machine-learning processes include projects and development at the Smithsonian Institution Archives, Harvard, and the State of Illinois and the University of Illinois.[142] The BitCurator Natural Language Processing project[143] addresses some of the additional issues around the collection and processing of huge numbers of heterogeneous digital records. The project applies existing open source tools to aid with the meaningful characterization of the raw text contents of files via entity relationships and topic modeling

Projects such as Citing and Archiving Research (CiTAR) and Reprozip[144] take computational techniques one step further in an effort to provide preservation and long-term access to software-based research resources, with Reprozip focused specifically on computational reproducibility for scientific research.

While some substantial progress has been made, work remains to be done, particularly around issues regarding personally identifiable information, security, and privacy concerns, and transparency and best practice of the stewardship community around these concerns.[145] Further work is also necessary in developing educational and continuing educational resources on computational techniques for both stewardship professionals and student communities, outreach and workflow documentation for a wide range of researchers and individuals that will need knowledge and understanding of both how to use these tools for preservation of their own digital materials or for access to materials they wish to use. Technological development also remains a large component of this work, including continued research into topics such as compression and restoration of preserved materials, description methods that leverage machine learning and natural language processing, examining the viability regarding the sustainability of emulating containers and environments, and more.

---

[142] See email project examples using a range of techniques from predictive coding to batch processing at the Smithsonian, Harvard University, and the University of Illinois: "Email Preservation - Collaborative Electronic Records Project," Smithsonian Institution Archives, accessed April 6, 2020, https://siarchives.si.edu/what-we-do/digital-curation/email-preservation-cerp; "Electronic Archiving System," Harvard University, accessed April 6, 2020, https://wiki.harvard.edu/confluence/pages/viewpage.action?pageId=194151171; and "Processing Capstone Email Using Predictive Coding," University of Illinois, accessed April 6, 2020, https://www.aits.uillinois.edu/services/professional_services/rims/about_rims/projects/processing_capstone_email_using_predictive_coding/.

[143] For further project details and technical information, see: "bitcurator/nlp," BitCurator GitHub, updated April 18, 2018, https://github.com/BitCurator/bitcurator-nlp/wiki.

[144] CITAR: Citing and Archiving Research, https://openpreservation.org/blogs/preserving-virtual-research-environments-introducing-citar-part-1/ "ReproZip," VIDA Group, NYU, accessed April 6, 2020, https://www.reprozip.org/.

[145] The New Media Consortium. *NMC Horizon Report: 2016 Museum Edition*, 2016. https://library.educause.edu/~/media/files/library/2016/1/2016hrmuseumEN.pdf; Micah Altman and Chris Bourg. "A Grand Challenges-Based Research Agenda for Scholarly Communication and Information Science," 2018. https://grandchallenges.pubpub.org/pub/final; Bergis Jules, "Confronting Our Failure of Care Around the Legacies of Marginalized People in the Archives," *Medium,* November 11, 2016, https://medium.com/on-archivy/confronting-our-failure-of-care-around-the-legacies-of-marginalized-people-in-the-archives-dc4180397280.

**Actionable Recommendations**

- Continue to support and scale the use of natural language processing and machine learning techniques in projects and in the development of workflows and practices in dealing with voluminous digital records.
- Integrate approaches to address information ethics and privacy into computational archival science methods and machine learning.
- Support stewardship community training, policy and workflow development documentation, and outreach around the application and sharing of computational techniques.

# 3. Organizational Policies and Practices

## 3.1 Advocate for Resources

Despite the economic upturn that has occurred since the first version of this report, many institutions, particularly small organizations, regularly struggle with obtaining enough resources for digital stewardship activities. This is evidenced by our survey of NDSA organizations, which list advocating for resources as a top priority (see [Appendix](#)), as well as the 2017 *NDSA Staffing Survey*,[146] in which respondents reported that the desired number of employees working on digital preservation activities was double what they currently had.

The Keepers Registry,[147] which combines the major efforts of CLOCKSS,[148] HathiTrust, the British Library, and the Library of Congress, among others, around the preservation of e-books and e-journals, provides one example of the importance of shared collection infrastructure and challenges of maintaining it. Launched in 2011, the effort is able to monitor the state of need for preservation of scholarly journals at a global scale, while continuing to build out search and discovery functionality that will aid in the identification of preservation gaps of journals in practice. However, as of July 31, 2019, the service was to be retired.[149] As of Fall 2019, the ISSN International Center and its Governing Board has agreed to maintain and keep the Registry's content available with plans to launch a new service in late 2019; however, this example underscores the fact that such critical resources are always at risk.

---

[146] NDSA Staffing Survey Working Group. "Staffing for Effective Digital Preservation 2017." https://osf.io/mbcxt/.
[147] "The Keepers Registry," ISSN International Centre and EDINA at the University of Edinburgh, accessed November 15, 2019, https://thekeepers.org/.
[148] CLOCKSS, accessed April 16, 2020, https://clockss.org/.

Even though fears of dramatic budget cuts to federal granting agencies have not been fully realized,[150] it is clear that reliance on external grant funding for ongoing stewardship activities is not sustainable. As the disappearance of substantial amounts of data from government websites with the start of the Trump administration demonstrates, relying exclusively on government agencies for preservation also carries significant risks.[151] Even in more favorable political climates, federal funding is often finite, as evidenced by NDSA's own change in organizational home from the Library of Congress to the Digital Library Federation due to the end of its funding program.[152]

As described in Section 2.2.4, many organizations are facing the heightened challenge of balancing priorities and allocating resources for digitization of high-risk analog collections (e.g. moving image and audio materials) with ongoing management of born-digital materials. The scope and severity of the risk to analog moving image materials has been well documented and discussed, but the development of a coordinated strategy amongst institutions is still needed to determine how best to utilize limited resources to address this challenge. In general, organizations must continue to advocate internally for appropriate resources, and appropriate reallocation of resources, to tackle the activities of digital stewardship. To make the case, stewarding organizations need to be able to offer value in exchange for the resources required to successfully address long-term digital stewardship issues. Resources such as the Digital Preservation Coalition's (DPC) Toolkit,[153] which helps practitioners craft a business argument for digital preservation, and their Executive

---

[149] "The Keepers Registry Funding Cessation Banner Announcement," April 25, 2019, http://web.archive.org/web/20190425225022/https://thekeepers.org/.

[150] Peggy McGlone. "For Third Year in a Row, Trump's Budget Plan Eliminates Arts, Public TV and Library Funding." *Washington Post*, March 18, 2019, https://www.washingtonpost.com/lifestyle/style/for-third-year-in-a-row-trumps-budget-plan-eliminates-arts-public-tv-and-library-funding/2019/03/18/e946db9a-49a2-11e9-9663-00ac73f49662_story.html.

[151] Harmon, Amy. "Activists Rush to Save Government Science Data—If they Can Find It." *New York Times*, March 6, 2017, https://www.nytimes.com/2017/03/06/science/donald-trump-data-rescue-science.html.

[152] "National Digital Stewardship Alliance," Library of Congress, accessed April 8, 2020, http://www.digitalpreservation.gov/ndsa/NDSAtoDLF.html

[153] "Digital Preservation Business Case Toolkit," Digital Preservation Coalition, last modified May 2, 2014, accessed April 8, 2020, http://wiki.dpconline.org/index.php?title=Digital_Preservation_Business_Case_Toolkit.

Guide[154] (done in partnership with UNESCO) help craft the case for funding preservation at the highest levels of leadership and government. The DPC also highlighted the current challenges and strategies for preserving social media[155] and transactional data[156] in Technology Watch Reports released in 2016, and the November 2018 revision of DPC's Bit List of Digitally Endangered Species[157] reiterated that the status of these materials remains vulnerable. Collaborative efforts such as these help link up advocacy and expertise across the preservation community.

**Actionable Recommendations**

- Continue to collaborate and amplify work done by the preservation community (e.g. the DPC's *Executive Guide*) that provides tools and communication strategies for making the case for preservation.
- Create and share preservation advocacy templates and business plans with the broader community that represent multiple sectors.
- Create crosswalks to advocacy templates and preservation services offerings.

## 3.2 Staffing and Training for Digital Stewardship

In 2017 the NDSA Standards and Practices Working Group released a *Staffing for Effective Digital Preservation 2017* report[158] that found that there was an increase in the percentage (46 percent, up from 34 percent in 2013) of respondents who were dissatisfied with "the way the digital preservation function was organized." The report authors note that additional questions are needed to further determine the details of why respondents are reporting increased levels of dissatisfaction. While more information and better understanding is needed, this increase potentially indicates that there is not one specific organizational approach to digital preservation activities (centralized in one department vs. distributed across multiple units) that is seen as more successful than others. On the training front, the 2017 report illustrated most respondents (68 percent) are still focusing on re-training existing staff, indicating that there continues to be a high need for training and continuing education programs for current professionals. Additionally respondents reported that training is one of the top benefits that they gain from participation in "consortia or cooperative efforts." More information on the details of these consortia-related training offerings would assist in better understanding their nature and content, as well as how they relate to other training programs. It should also be noted that most (46 percent) respondents of the 2017 survey were from academic institutions, which was about the same (45 percent) in the 2012 survey results. This is not necessarily surprising, but demonstrates the need for additional outreach and effort to reach a broader type of institution in future staffing surveys. In particular, smaller organizations such as public libraries, museums, and historical societies are likely to have different needs when it comes to digital preservation staffing and training. Without an increase in responses from people working at these kinds of institutions, the data needed to better understand these needs will not be collected.

---

[154] "Executive Guide on Digital Preservation," Digital Preservation Coalition, accessed April 8, 2020, https://en.unesco.org/news/unesco-and-dpc-release-executive-guide-digital-preservation.

[155] Sara Day Thomson. "Preserving Social Media," February 2016, https://www.dpconline.org/docs/technology-watch-reports/1486-twr16-01/file.

[156] Sara Day Thomson. "Preserving Transactional Data," May 2016, https://www.dpconline.org/docs/technology-watch-reports/1525-twr16-02/file.

[157] Digital Preservation Coalition, "Bit List of Digitally Endangered Species," 2018, https://www.dpconline.org/docs/miscellaneous/advocacy/1932-bitlist2018-final/file

[158] NDSA Staffing Survey Working Group. "Staffing for Effective Digital Preservation 2017," 2017, https://osf.io/mbcxt/.

Since its pilot program in 2013, the National Digital Stewardship Residency (NDSR)[159] program has proven to be a very successful training model, placing recent master's degree graduates with academic, federal, non-profit, or cultural heritage organizations to work on a digital stewardship project.[160] Residents have an immediate opportunity to implement their skills and knowledge and make a practical significant impact in advancing digital stewardship activities in a variety of institutions. Perhaps the most successful aspect of the program, as indicated by resident feedback, is the use of the cohort model.[161] Residents pointed to the benefit of opportunities for regular communication and support from the group of individuals that began their residencies together. The strong connections formed amongst cohort groups have provided an ongoing network of support as these individuals have launched their professional careers post-NDSR residency. Now in its sixth year, the program has continued to evolve, establishing governance in the form of an advisory board, producing valuable resources like the NDSR Handbook and Toolkit, and articulating challenges such as the need for addressing sustainability beyond its initial grant-funded period. Residents have also made their own contributions to documenting the successful

---

[159] "NDSR: National Digital Stewardship Residency," accessed April 17, 2020, https://ndsr-program.org/.
[160] Meridith Beck Mink. "Keepers of our Digital Future, An Assessment of the National Digital Stewardship Residencies, 2013-2016," December 2016, https://www.clir.org/pubs/reports/pub173/.
[161] The National Digital Stewardship Residency. "The National Digital Stewardship Residency Handbook and Toolkit Version 1.0," 2017. https://ndsrprogram.files.wordpress.com/2017/10/ndsr_handbook_and_tooklit2.pdf.

aspects of the program[162] as well as outlining the competencies required for digital stewardship.[163]

**Actionable Recommendations**

- Explore and test models for sustainability of digital preservation training programs such as the NDSR.
- Conduct a longitudinal analysis of the effectiveness of digital preservation education and training programs to determine the best approaches to meeting digital preservation staffing needs.
- Engage the community in ongoing dialogue around central efforts, such as the Levels of Preservation reboot, that will have an education, training, and advocacy component.

# 4. Technical Infrastructure Development

## 4.1 Coordinating an Ecosystem of Sustainable Shared Services

Although the digital stewardship community continues to make great strides in identifying the existing gaps in a modular, community-wide digital stewardship infrastructure and developing tools and services to fill those gaps, coordinating an ecosystem of sustainable shared services remains a significant challenge. The 2015 *National Agenda* identified the widespread reliance on project- or grant-based funding and a lack of effective coordination in ensuring that community-developed tools are sustainably maintained over time as significant challenges for digital stewardship. Following on that finding, the 2017-2018 survey of key decision makers for long-term access planning at NDSA member organizations and other cultural heritage institutions (described in Section 1.1) found that "maintaining ongoing integration, interoperability, and collaborative projects" was indicated as one of the highest priorities—and most difficult challenges—to be tackled.

It is within this context that the Digital Preservation Network (DPN) in December 2018 announced its decision to wind down and sunset its operations. In her announcement, DPN Executive Director Mary Molinaro cited a lack of economic sustainability in DPN's financial model as a major factor in its demise. In order to maintain its digital storage services, "a critical mass of member institutions" would need to make large storage deposits.[164] However, membership numbers declined substantially, especially in the summer of 2018; at the same time, over half of the member institutions never made a single deposit, and very few members ever purchased additional storage beyond the 5 TB allocated annually to each member. Although DPN was not able to sustain its operations, its initial work that funded infrastructure development at its constituent partner nodes (APTrust, Chronopolis, HathiTrust, and the Texas Digital Library) will continue to be repurposed to support long-term preservation and access for members of those organizations. A group (composed, in part, of former DPN nodes) has, in fact, continued to meet and explore possible collaborative service models under the name of the Distributed Digital Preservation Services Collaborative.[165]

---

[162] Rebecca Fraimow, Meridith Beck Mink, and Margo Padilla. "The National Digital Stewardship Residency: Building a Community of Practice through Postgraduate Training and Education." *D-Lib Magazine*, Volume 23, Number 5/6 (May/June 2017), https://doi.org/10.1045/may2017-fraimow.
[163] Karl-Rainer Blumenthal et al., "What Makes a Digital Steward: A Competency Profile Based on the National Digital Stewardship Residencies," LIS Scholarship Archive (14 July 2017), doi:10.31229/osf.io/tnmra.

In his analysis of DPN's closure,[166] Roger Schonfeld points out several factors that could be applicable to the continued operation of other collaborative and shared services. For instance, "a clear product offering took time to emerge," leading to a lack of understanding of the value of DPN membership in an environment in which many universities are consolidating their enterprise storage solutions in less costly cloud providers. Along the same lines, Schonfeld asserts that "membership models are ill-suited to product organizations and marketplace competition;" in order for a service provider like DPN to remain viable, it needs to deliver a clear solution to a well-defined problem for a reasonable price—an issue which other membership-based organizations in the digital stewardship and cultural heritage sector will certainly need to consider in the coming years. In fact, organizations such as the BitCurator Consortium and the Software Preservation Network (SPN), originally created through grant funding, have been vocal and transparent about issues around long-term sustainability. SPN Community Manager Jessica Meyerson reiterated the organization's commitment to forming a healthy, secure community with a strong and well-informed governance and business infrastructure.[167] Ultimately, DPN's closure may serve as the impetus for other groups and organizations to reexamine their product offerings and business models, and we may see further consolidation and reorganization—such as the merger between Lyrasis and Duraspace[168]—in this arena in the future.

---

[164] "Information Update," The Digital Preservation Network, December 5, 2018, accesssed April 27, 2020, https://web.archive.org/web/20190226161550/http://dpn.org/news/2018-12-05-information-update.
[165] dps collaborative, "Digital Preservation Services Collaborative," accessed April 27, 2020, https://dpscollaborative.org.
[166] Roger C. Schonfeld. "Why Is the Digital Preservation Network Disbanding?" *The Scholarly Kitchen* (blog), December 13, 2018, https://scholarlykitchen.sspnet.org/2018/12/13/digital-preservation-network-disband/.

These issues are also a concern for academic libraries' roles and responsibilities around scholarly communications. The *NMC Horizon Report: 2017 Library Edition*[169] described ongoing integration, interoperability, and collaborative projects as a "difficult challenge"—challenges that we understand but for which solutions are elusive. Increasingly, as academic libraries take on a greater role in the research data lifecycle, an emphasis is emerging in scholarly communication on moving from a single system-based mindset for housing research outputs to a broader range of interoperable systems and service offerings, including data management, analysis, and preservation. For instance, within a single academic or research library, there may be separate yet overlapping digital repositories, research information management systems, journal publishing platforms, and more, and these systems must be harmonized to operate together in order to improve the ease with which librarians can manage the variety of data and share it with researchers, funders, and other stakeholders.

In recent years, institutions have sought to address these scholarly communication issues by developing the underlying technical framework to support interoperability between their systems, using APIs and other well-documented standards and protocol for harvesting and sharing metadata and other content. OAI-PMH[170] has been broadly incorporated into repository software and systems to encourage a standardized format for information exchange, and ORCID iD[171] has emerged as a standardized form of identification for researchers across all disciplines.

---

[167] Jessica Meyerson, "Community Cultivation and the Software Preservation Network," November 30, 2018, accessed April 29, 2020, https://www.dpconline.org/blog/idpd/community-cultivation-and-spn.

[168] "Amplifying Impact: LYRASIS and DuraSpace Announce Intent to Merge," LYRASIS, posted January 23, 2019, accessed April 8, 2020, https://duraspace.org/amplifying-impact-lyrasis-and-duraspace-announce-intent-to-merge-2/.

[169] The New Media Consortium. "NMC Horizon Report: 2017 Library Edition," 2017, https://library.educause.edu/~/media/files/library/2017/12/2017nmchorizonreportlibraryEN.pdf.

[170] "Open Archives Initiative Protocol for Metadata Harvesting," Open Archives Initiative, accessed April 16, 2020, https://www.openarchives.org/pmh/.

[171] "ORCid," ORCID, accessed April 16, 2020, https://orcid.org/.

In the realm of cultural heritage content, the rapid expansion of collections such as moving image, web archives, and other large born-digital and digitized collections has revealed pain points for managing extremely large collections in a distributed service environment. For instance, as discussed in Section 2.2.4, the resolution and file size of born-digital audiovisual media is rapidly increasing, and these large media files take more time to move, copy, process, and store. Long-term preservation and access also necessitates migration to newer digital storage as storage media and formats become obsolete; at the same time, the increased reliance by many academic and cultural heritage institutions on distributed and cloud storage for backing up or replicating large digital collections could, in some cases, also increases the complexity of migration and management of these digital files.

Cloud computing has emerged as a crucial component in the ecosystem of shared services for digital preservation. Broadly speaking, in the last several years cloud computing services have allowed for greatly increased flexibility and affordability in operationalizing a digital preservation program and building out the required technical infrastructure.[172] In addition, these services can help academic and cultural heritage institutions achieve digital preservation recommendations and best practices, including replication of content in geographically distributed locations and regular fixity checks. Many small and mid-sized institutions, or those with small budgets for digital preservation, may be tempted to store their content with a super-low-cost cloud storage provider, such as Amazon's Glacier cold/near-line storage. However, as early as 2012, digital preservation professionals began to raise concerns about Glacier[173]—Amazon's durability and reliability claims are largely untested, their terms of service and other customer agreements are opaque, and while Glacier storage may seem cheap initially, getting data back from Amazon is not. Many services are designed to make deposit easy and cheap; costs are incurred when a depositor needs to take content back out. The latest version of the Digital Preservation Storage Criteria,[174] released in 2018, includes categories such as content integrity, information security, scalability and performance, and transparency. The criteria document can be used to help compare services and service providers, inform discussions within an institution, or prioritize content for different levels of management.

---

[172] "Digital Preservation Handbook: Cloud services," Digital Preservation Coalition, accessed April 8, 2020, https://www.dpconline.org/handbook/technical-solutions-and-tools/cloud-services.

[173] "Amazon's Creeping 'Glacier' and Digital Preservation," Program on Information Science, MIT Libraries, posted November 15, 2012, accessed April 8, 2020, https://web.archive.org/web/20190612002607/https://informatics.mit.edu/blog/2012/11/amazon%E2%80%99s-creeping-%E2%80%98glacier%E2%80%99-and-digital-preservation.

[174] "Digital Preservation Storage Criteria," January 2018, https://osf.io/sjc6u/.

Despite the challenges outlined above, recent institutional collaborations represent an important step in the right direction. National and regional repository networks, such as the Confederation of Open Access Repositories (COAR) and the Digital Repository Federation,[175] have emerged to facilitate the strong partnerships and fora for communication that are required to successfully increase integration, interoperability, and collaboration between research institutions. Similarly, in 2017-2018, a group of nonprofit digital preservation service providers, including APTrust, Chronopolis, DuraSpace, and Educopia/MetaArchive Cooperative,[176] collaborated to produce the Digital Preservation Declaration of Shared Values[177] as a foundation for future efforts, including potential interoperability amongst service providers.

In addition to these emerging multi-institutional collaborations and initiatives that are more top-down in nature, there have also been a number of recent collaborative projects that have produced valuable resources with the potential to significantly impact digital preservation practice. These efforts have been informal and ad-hoc in nature, driven by motivated practitioners coming together to identify shared needs and then moving forward to produce the resources to address those needs. Typically taking place outside of formal grant-funded projects or standards bodies, these projects illustrate how gaps between high level digital preservation standards and daily practice are being filled by passionate individuals. Harnessing and connecting these efforts through more structured frameworks could result in more alignment and strengthening of ties between institutions in the digital stewardship community towards achieving collective goals. One of the key roles served by the NDSA is to identify and foster such activity to provide longitudinal alignment. For example, the Levels of Preservation Reboot Working group, established in 2018, completed the second version of the Levels of Preservation in the fall of 2019, is an example of such a global collaboration. In addition, work currently underway by various distributed services aimed at providing a shared lexicon of service options across the preservation service landscape will likely become part of the overall NDSA objective efforts towards preservation transparency and assurance. Educopia's Community Cultivation Field Guide[178] is another excellent source for the community.

---

[175] The New Media Consortium. "NMC Horizon Report: 2017 Library Edition," 2017, https://library.educause.edu/~/media/files/library/2017/12/2017nmchorizonreportlibraryEN.pdf.

[176] "MetaArchive," accessed April 16, 2020, https://metaarchive.org/.

[177] "Digital Preservation Declaration of Shared Values: Comments and Interest Welcome," LYRASIS, March 26, 2018, https://duraspace.org/digital-preservation-declaration-of-shared-values-comments-and-interest-welcome/.

**Actionable Recommendations**

- Continue to develop a matrix of services across the preservation landscape that encourages both transparency and awareness of the roles each service plays.
- Foster and refine the Levels of Preservation on a regular basis to keep pace with changes in digital preservation.

## 4.2 Emerging File Formats and Strategies

In the previous edition of the *National Agenda*, recommendations encouraged digital stewardship organizations to document the file formats they are currently managing and to develop action plans outlining how they plan to monitor and manage their digital file formats long-term, prioritizing those formats that are at a high risk of obsolescence. Publicly available documentation on file formats and format action plans, such as the Library of Congress's Sustainability of Digital Formats website[179] and the National Archives and Records Administration's Bulletin 2014-04 Revised Format Guidance for the Transfer of Permanent Electronic Records,[180] provides a comprehensive outline and reference point on a wide variety of file formats currently managed by stewardship institutions and encourages institutions to coalesce around a smaller set of preferred digital format options for long-term preservation and access. This foundational documentation work has allowed the development of tools and software that implement and manage abstract format policies and action plans. For example, the Archivematica digital preservation platform and its format policy registry[181] contains a list of file formats recognized by the software along with user-configurable scripts that automate preservation actions specific to each file format, including file identification, characterization, extraction, and normalization. And in 2018, representatives from Artefactual (the developers behind Archivematica), Arkivum, Preservica, and Jisc first presented on the concept of Preservation Action Registries, or PAR, which would provide machine-readable preservation action recommendations through APIs, allowing these recommendations to be shared among digital preservation platforms or peer institutions in a more interoperable way.[182]

---

[178] Katherine Skinner. "Community Cultivation: A Field Guide," November 7, 2018, https://educopia.org/cultivation/.

[179] "Sustainability of Digital Formats: Planning for Library of Congress Collections," Library of Congress, updated April 23, 2019, accessed April 9, 2020, https://www.loc.gov/preservation/digital/formats/index.html.

[180] "Bulletin 2014-04, Format Guidance for the Transfer of Permanent Electronic Records," National Archives, revised August 2018, accessed April 9, 2020, https://www.archives.gov/records-mgmt/bulletins/2014/2014-04.html.

[181] "Format policy registry requirements," Archivematica, revised March 23, 2017, accessed April 9, 2020, https://wiki.archivematica.org/Format_policy_registry_requirements.

While these developments represent a significant step forward in efforts to manage expanding digital content collections in an automated way, emerging file formats and content types present new challenges for the institutions charged with harvesting, managing, and providing access to cultural heritage data. As noted in Section 2, content types such as scientific data sets, social media, and dynamic web content are increasingly recognized as a vital part of the public record, yet they are at a high risk of data loss.

Another increasingly ubiquitous stream of content that presents challenges for harvest and preservation is documents or other digital objects that are created and modified within a web browser, outside of the traditional file format/software environment. For instance, documents created using Google Drive apps are not stored, shared, or edited in that environment as a single preservable file, enabling interactive/dynamic features such as collaborative sharing and editing, commenting and notifications, and group chat. Documents created natively in Google services can be exported in various file formats, including preservation-worthy non-proprietary formats such as OpenDocument Format (*.odt), but potentially significant embedded metadata, including creation/last modified dates, timestamps on comments, and other revision details, is lost on export.[183] For the time being, archivists and other practitioners may be tasked with assessing documents for their informational and research value in order to determine which of these additional features or data points, if any, is significant for long-term access and preservation; this could guide the selection of an export file format.[184]

---

[182] Matthew Addis, Justin Simpson, Jonathan Tilbury, Jack O'Sullivan, and Paul Stokes. "Digital Preservation Interoperability through Preservation Actions Registries." Paper presented at iPres 2018, Boston, MA, September 25, 2018, https://figshare.com/articles/Digital_Preservation_Interoperability_through_Preservation_Actions_Registries/6628418.

[183] "Tools: Google Takeout," University of Wisconsin-Madison Research Data Services, posted August 22, 2013, accessed April 9, 2020, http://researchdata.wisc.edu/tag/google/.

[184] Jenny Mitcham. "How can we preserve Google documents?" *Digital Archiving at the University of York* (blog), posted April 28, 2017, accessed April 9, 2020, http://digital-archiving.blogspot.com/2017/04/how-can-we-preserve-google-documents.html.

Application Programming Interfaces (APIs) provide the technological infrastructure for interacting with and harvesting data from web-based applications such as social media platforms. Collecting institutions can freely acquire raw data from some sources or platforms using APIs, either using local software development expertise or using a third-party archiving service. Other platforms provide access to current and/or historical data only through data resellers, such as Gnip for historical Twitter data or DataSift for Facebook topic data.[185] No matter which technological method or infrastructure is used for harvesting dynamic web content, the fact remains that the institutions charged with stewarding these materials must proactively seek out and develop relationships with content creators in order to capture web content soon after its creation, otherwise it will be vulnerable to modification or data loss. Furthermore, deliberate preservation planning to identify and select strategies for ensuring long-term preservation and access to this type of content must also occur early in the process so that content is captured and handled in a way that ensures it will be relevant and useful to future users.[186]

Finally, emulation has been proposed as a strategy for preservation and access to obsolete digital media since at least the late 1990s,[187] but until recently very few tools and services have been made widely available to stewardship organizations who may want to implement emulation of applications, operating systems, or hardware platforms to ensure that the "look and feel" and interactivity of digital objects is maintained long-term. A 2017 Sloan Foundation grant is supporting efforts at Yale and the Software Preservation Network to further develop an open source software tool called bwFLA that enables the creation, management, and distribution of "virtual machines" that can simulate the hardware of an older computer on a newer computer and then run older software on the simulated machine.[188] As mentioned previously in Section 2.2.3, the Software Preservation Network[189] and Yale University Library's "Scaling Emulation and Software Preservation Infrastructure" Andrew W. Mellon grant project is creating a scalable, community-driven

---

[185] Sara Day Thomson. "Preserving Social Media," February 1, 2016, https://www.dpconline.org/docs/technology-watch-reports/1486-twr16-01/file.

[186] Sara Day Thomson. "Preserving Transactional Data," May 2016, https://www.dpconline.org/docs/technology-watch-reports/1525-twr16-02/file.

[187] Stewart Granger, "Emulation as a Digital Preservation Strategy," *D-Lib Magazine* Volume 6 Number 10 (October 2000), http://www.dlib.org/dlib/october00/granger/10granger.html.

[188] Alfred P. Sloan Foundation Grants, Yale University emulation and software preservation grant, 2017, https://sloan.org/grant-detail/8228.

[189] "Software Preservation Network," accessed April 1, 2020, http://www.softwarepreservationnetwork.org/.

Emulation as a Service Infrastructure (EaaSI)[190] that will allow institutions to share software and build their capacity for emulation as a digital preservation strategy. The distributed management and community engagement elements of the EaaSI project represent an important move forward in centralizing our efforts and reducing redundant work across institutions, so that emulation may finally become more widely adoptable as a technological strategy for preservation and long-term access.

**Actionable Recommendations**

- Support emerging digital preservation formats and strategies via community efforts that aim to develop scalable, flexible technical and administrative infrastructures.
- Research, develop, and share digital preservation policies and workflows related to web applications, APIs, and cloud-based digital materials.
- Increase community contributions to existing file format identification and deprecation identification efforts (such as PRONOM,[191] bit-list review, standards review, etc.) to allow for community-wide adoption of best practices and migration/emulation at scale.

## 4.3 Integration Across Digital Preservation Providers and Systems

In the years since the previous edition of the *National Agenda* was released, there have been significant changes in the landscape of digital preservation providers and systems. As noted in Section 4.1, the dissolution of DPN in late 2018 called into question the economic sustainability of traditional notions of long-term preservation and storage. At the same time, many other cloud-based storage options, both in the cultural heritage community and in the commercial sector, have emerged as viable and less costly options for storing at least a subset of the multiple copies that institutions pursuing more than one storage strategy will manage. With this proliferation of options for distributed digital preservation, the need for integration across digital preservation providers and systems—and interoperability between them—becomes even more vital.

---

[190] "About EaaSI," Software Preservation Network, accessed April 9, 2020, http://www.softwarepreservationnetwork.org/eaasi/.

[191] "PRONOM," The National Archives (UK), accessed April 16, 2020, https://www.nationalarchives.gov.uk/PRONOM/Default.aspx.

The *Beyond the Repository: Integrating Local Preservation Systems with National Distribution Services* report,[192] released in 2018, provides a comprehensive overview of the issues facing digital preservation practitioners in a distributed storage environment. In an IMLS-funded project, Northwestern University Libraries and the University of California San Diego Library investigated "how local digital preservation practices and repository systems interoperate with distributed digital preservation (DDP) services," such as The Digital Preservation Network (DPN), Chronopolis, and the APTrust, through a survey on institutions' digital preservation programs and in-depth follow-up interviews with selected survey participants. Survey responses revealed that most respondents are storing two or three copies of their content in different locations, but some institutions are keeping seven or more. Most respondents are also pursuing more than one storage strategy, storing copies in a variety of locations including multiple locations onsite, commercial cloud storage, and distributed digital preservation providers. Almost half of the survey respondents indicated that they send only a subset of their content to offsite storage such as a DDP provider or cloud service, indicating that curatorial decisions and prioritization are an integral part of their digital preservation practices.

Within this context, digital preservation practitioners face several challenges. Chief among these is the lack of interoperability between the various software tools and repository systems used locally for file management, and between locally stored content and offsite storage providers. Furthermore, participants commonly reported difficulties in tracking their content between systems. Some survey participants and interviewees "described their systems as separate units with little integration between them, requiring manual processes and workarounds."[193] One notable initiative attempting to address this issue is the OSSArcFlow[194] project, which is investigating and modeling a range of workflows for born-digital archival content that incorporate three leading open source software platforms—BitCurator, Archivematica, and ArchivesSpace. By documenting current practices at twelve partner institutions, supporting development and testing of aspirational workflows, and publishing an implementation guide based on the findings, the OSSArcFlow project aims to help digital preservation practitioners more easily automate the movement of content between software tools.

---

[192] Evviva Weinraub et al., "Beyond the Repository: Integrating Local Preservation Systems with National Distribution Services," 2018, accessed April 9, 2020, https://doi.org/10.21985/N28M2Z.
[193] ibid., page 23.
[194] "OSSArcFlow," Educopia Institute, 2017-2019, accessed April 9, 2020, https://educopia.org/ossarcflow/.

Unsurprisingly, lack of required funding or staffing for a robust digital preservation program was cited by many Beyond the Repository survey participants as one of the primary organizational or institutional barriers to the adoption of DDP services for preservation storage, as well as the adoption of digital preservation policies more broadly.[195] The decreasing cost of commercial cloud storage, such as Amazon Glacier or Deep Archive, in addition to the availability of more robust storage options such as Chronopolis or APTrust, makes a tiered approach to keeping multiple offsite copies (based on criteria such as institutional mandates or importance and uniqueness of content) more viable. However, for such a tiered approach to be able to be broadly implemented among cultural heritage institutions, better methods of automatically tracking and managing multiple copies of content across providers and systems must be developed.

Another promising initiative in the area of interoperability is the Oxford Common File Layout (OCFL).[196] This specification describes an application-independent approach to the storage of digital information in a structured, transparent, and predictable manner. It is designed to promote long-term object management best practices within digital repositories. Once content has been accessioned to a digital repository, it is unlikely to change significantly over its lifetime; by contrast, the software and systems used to manage this content are in constant flux, requiring continual updates and migrations to new systems. By providing a specification for the file and directory layout on disk, OCFL aims to reduce or eliminate the need for transitions between application-specific methods of file management.

Following on the work of the Beyond the Repository report described above, the One to Many: Preserving Local Repository Content in Distributed Digital Preservation Systems[197] Andrew W. Mellon-funded grant managed by UC San Diego aims to address the challenges in moving and syncing content between institutions' local repositories and their distributed digital preservation storage. The project will integrate Samvera repository software with

---

[195] Evviva Weinraub et al., "Beyond the Repository: Integrating Local Preservation Systems with National Distribution Services," 2018, accessed April 9, 2020. https://doi.org/10.21985/N28M2Z.

[196] Oxford Common File Layout Specification 0.1, October 18, 2018, accessed April 9, 2020, https://ocfl.io/0.1/spec/.

[197] UC San Diego News Center, "UC San Diego Library Receives Mellon Grant to Develop Approaches to Preserving Digital Repositories," February 5, 2019, https://ucsdnews.ucsd.edu/pressrelease/uc_san_diego_library_receives_mellon_grant_to_develop_approaches_to_preserving_digital_repositories.

Chronopolis preservation storage and may use OCFL as its versioning system, allowing preservationists to more easily manage content in a secure, standards-based way.

**Actionable Recommendations**

- Encourage digital repository systems and distributed digital preservation services to more broadly implement standards and best practices, such as the shared BagIt profile, to ensure the mobility of content between systems.
- Building on current grant-funded work, develop and make more widely available methods for managing multiple copies of content stored locally and in distributed providers, including support for versioning files and/or metadata, running and documenting fixity checks over time, etc.
- Promote the creation of repository systems based on and conforming to the Oxford Common File Layout (or a future structured storage specification, if it becomes more widely adopted).

# 5. Research Priorities

## 5.1 Strengthening the Evidence Base for Digital Preservation

A common challenge running through this report, and an overarching challenge for research priorities, is the limited amount of empirical evidence available on preservation—and the relative dearth of testbeds, common corpora, longitudinal tracking surveys, reliable computer simulations, and rigorous scientific test designs for preservation research. For example, this report makes clear that effective digital preservation relies on answering questions such as: What content is already being effectively stewarded by other organizations? How much is the expected future cost of preserving that content? What is the likelihood that a community will use a collection in the future, and how will they use it? How do we predict the likelihood of future preservation threats? What is the reliability of current digital preservation services and organizations? And how successful are other proposed strategies for replication, monitoring, certification, and auditing at preventing loss due to these threats?

These questions are not new—they were raised in the prior *National Agenda*, five years ago—but they remain substantially unanswered and more important than ever. The continuing importance of these questions is underlined not only by the gravity of the applied challenges described in earlier sections of this report, but also by the potential for advances in digital preservation research to support a more open, equitable, inclusive, and sustainable future for the information ecosystem. The recently published 2018 MIT "Grand Challenges-Based Research Agenda for Scholarly Communication and Information

Science"[198] emphasizes that durability of digital information is a critical challenge to world knowledge, and to our future.

Over the last five years, moderate to larger efforts have been undertaken with an increase in the amounts of funding allocated to research and development of digital preservation policies, planning, and practices, with funding from many federal and foundation funders. There have been a number of recent awards aimed at furthering the state of knowledge, implementation, education, and staffing of digital preservation efforts. A prime example is the systematic approach that the IMLS has taken in articulating and supporting a coordinated platform of research and practice.[199] While many of these projects do not have readily measurable outputs as of yet, they have involved curators more deeply in the research process, and the efforts in this direction signal positive potentials for digital preservation research and development.

Longitudinal surveys of community practice[200] continue to provide an important part of the evidence base. Looking across these surveys demonstrates the rapid growth in stewarded content, the incremental advances in practices and resources, and the substantial gaps in organizational preservation practice, resourcing, and planning.

Building a common evidence base around the disposition of content stewarded across the community has proved harder. To illustrate, consider two substantial initiatives in the community: the Digital Preservation Network (DPN), and The Keepers Registry, that aimed to provide a broader evidence base but faced significant challenges (these are discussed in more detail in Sections [3.1](#) and [4.1](#)). DPN, although it was billed primarily as a preservation service, aimed to establish transparency around the durability of community content through open succession planning and reporting. As discussed in more detail in prior sections of this report, DPN was not successful in establishing a sustainable cost model—in part because individual institutions were not willing to pay a premium for DPN's storage service. In addition, DPN's services lacked a clear content strategy, compounded by a commensurate lack of transparency to its own community . The Keepers Registry, discussed in Section 3.1, had been more successful, but did not obtain continued funding. Moreover, while major stewardship organizations plan to expand sharing information on and access to the collections that they preserve[201] (the Library of Congress, for example, has made it a core priority)[202], these collections do not provide a comprehensive evidence base for research on preservation, since they do not expose the preservation practices in use for these collections, nor their outcomes. Overall the evidence base of digital preservation remains substantially incomplete.[203]

---

[198] Micah Altman and Chris Bourg. "A Grand Challenges-Based Research Agenda for Scholarly Communication and Information Science," 2018, https://grandchallenges.pubpub.org/pub/final.

[199] Trevor Owens et al. "NDP at Three: The First Three Years of IMLS Investments to Enhance the National Digital Platform for Libraries," 2017, https://www.imls.gov/sites/default/files//publications/documents/imls-ndp-three-508.pdf; Trevor Owens et al., "Digital Infrastructures that Embody Library Principles: The IMLS National Digital Platform as a Framework for Digital Library Tools and Services." In *Applying Library Values to Emerging Technology: Tips and Techniques for Advancing within Your Mission,* #72. Publications in Librarianship. ACRL, 2018, http://www.ala.org/acrl/sites/ala.org.acrl/files/content/publications/booksanddigitalresources/digital/9780838989401.pdf.

[200] NDSA Community surveys including NDSA Web Archiving survey report, now published for the fourth time in 2018, the NDSA Staffing for Effective Preservation report, published for the second time in 2017 and the NDSA Storage survey now published for the fourth time (collectively) comprise an important part of this evidence base. Find all the NDSA reports at https://osf.io/4d567/.

Five years ago, we noted that medium-scale observational studies and field experiments had provided useful insights into the failure rates of spinning disk storage.[204] In the intervening time, techniques for enhancing the short-term reliability of digital storage have advanced steadily,[205] but there remains a paucity of systematic empirical evidence on the long-term reliability or sustainability of content or failures *in vivo*. Moreover, modeling and data collection on the institutional factors that are responsible for long-term preservation failures has progressed little. As the selections below summarize, over this period, there has been only incremental progress in closing the gap between the scale of information production and preservation; in developing trust models at all levels; and in developing models of future value and cost—including the environmental costs of long-term digital preservation.

During the last five years, technological advances in production, storage, and transformation of content, and changes in the economics and organization of information production have continued to outpace the understanding of how to meaningfully preserve that content. For example, see analyses of the changing landscapes of journalism, art and

---

[201] J. Stephen Downie et al., "The HathiTrust Research Center: Exploring the Full-Text Frontier," *Educause Review* (May/June 2016)*;* Maja Kominko, ed. *From Dust to Digital: Ten Years of the Endangered Archives Programme*. Cambridge, UK: Open Book Publishers, 2015, https://doi.org/10.11647/OBP.0052.

[202] "Digital Strategy for the Library of Congress," Library of Congress, accessed April 9, 2020, https://www.loc.gov/digital-strategy.

[203] Jeremy York, Myron Gutmann, and Francine Berman, "What Do We Know About the Stewardship Gap," *Data Science Journal* 17 (17 August 2018): 19, http://doi.org/10.5334/dsj-2018-019.

[204] Eduardo Pinheiro, Wolf-Dietrich Weber, and Luiz André Barroso, "Failure trends in a large disk drive population." In *Proceedings of 5th USENIX Conference on File and Storage Technologies*, 2007, https://research.google/pubs/pub32774/.

[205] Rekha Nachiappan et al., "Cloud storage reliability for big data applications: A state of the art survey," *Journal of Network and Computer Applications* 97 (2017): 35-47.

culture, websites, algorithmically produced content, research data, and software.[206]

Almost ten thousand scholarly publications with a relation to digital-preservation research are produced each year[207]—a number comparable to publications on blockchain—and such research has been produced at a steady pace over the last five years. However, much of this work continues to focus on isolated case studies, and more generally digital preservation research lacks the level of integration and cross-comparability of research in, for example, cryptographic methods or storage technologies. Generally, case studies remain over-represented in digital preservation research, and articles in the field are less likely to contain highly-cited work than works in related computer- and information-science fields.

To systematically guide decisions in this area, case studies must be repeated longitudinally, repeated in different environments, and transformed, eventually, into production public testbeds[208] and conformance tests that can be used to rigorously compare approaches and systems. Furthermore, the research community still lacks shared, durable, community testbeds that provide a place where tools can be tested and a common set of digital content with which to run trials.This practice could provide a solution for systematically comparing, proposing, and incrementally improving practice, calibrating both theory models and practical understanding in the process.

Moreover, a search of the discipline's key reference works, bibliographies, and literature databases[209] reveal very few rigorously validated preservation methods, wide-scale

---

[206] See Section 2.2, and, Oya Y. Rieger et al. "Preserving and emulating digital art objects," November 2015, https://ecommons.cornell.edu/handle/1813/41368.; Mohamed Aturban, Michael L. Nelson, and Michele C. Weigle, "Difficulties of Timestamping Archived Web Pages," December 8, 2017, https://arxiv.org/pdf/1712.03140.pdf; Sharon Ringel and Angela Woodall. "A Public Record at Risk: The Dire State of News Archiving in the Digital Age," March 28, 2019, accessed April 9, 2020, https://www.cjr.org/tow_center_reports/the-dire-state-of-news-archiving-in-the-digital-age.php.; Clifford Lynch, "Stewardship in the 'Age of Algorithms'," First Monday 22, no. 12 (2017, https://firstmonday.org/article/view/8097/6583.

[207] This approximation is based on keyword searches using google scholar across date ranges from 2014-2018.

[208] Recent research by Becker, Faria, & Duretec, as part of the BenchmarkDP project, provides a potential model based framework for such testbeds. See: Christoph Becker, Luis Faria and Kresimir Duretec, "Scalable Decision Support for Digital Preservation: An Assessment," *OCLC Systems & Services: International Digital Library Perspectives*, Emerald Publishing, 2015; Cristoph Becker and Kresimir Duretec, "Free benchmark corpora for preservation experiments: using model-driven engineering to generate data sets." In *Proceedings of the 13th ACM/IEEE-CS joint conference on Digital libraries*. ACM, 2013.

[209] Uwe M. Borghoff et al., *Long Term Preservation of Digital Documents* (Cham, Switzerland: Springer, 2005).; David Giaretta, *Advanced Digital Preservation* (Berlin: Springer-Verlag, 2011).; "Curation

empirical studies, probability-based surveys or field experiments, replicable simulation experiments, public test corpuses, testbeds,[210] or recognized conformance tests.[211] Although an applied field cannot rely on theoretical literature alone, it is essential to both grounded theory and robust practice that preservation strategies, methods, tools, and measures be formalized, standardized and evaluated systematically and rigorously. Broadly, across the field of digital preservation, there is an urgent need to develop a modular open and robust approach to testing, conformance, and measurement,[212] in order to extend the evidence base on which preservation research and policy is founded.

Evidence is needed to support both general selection of digital preservation practices and methods, and applications of selected digital preservation methods in a specific operational context. What is also needed is to apply the research methodologies already used in other fields that rely heavily on observation of human and system behavior. This includes methodologies such as: probability-based surveys of information management practice and outcomes; replicable simulation experiments, and theoretically grounded new practices, tools, and methods; and field experiments, in which randomized interventions are applied and evaluated in real operational environments.

**Actionable Recommendations**

- Funders should give priority to programs that systematically contribute to the overall cumulative evidence base for digital preservation practice and resulting outcomes—including supporting testbeds for systematic comparison of preservation practices.

---

Manual," Digital Curation Centre, 2012, accessed April 9, 2020, http://www.dcc.ac.uk/resources/curation-reference-manual.; Charles W. Bailey Jr., "Digital Curation Bibliography: Preservation and Stewardship of Scholarly Works," *Digital Scholarship,* 2012, http://digital-scholarship.org/dcbw/dcb.htm.; Blue Ribbon Task Force on Sustainable Digital Preservation and Access. "Sustainable Economics for a Digital Planet: Ensuring Long-Term Access to Digital Information," February 2010, http://brtf.sdsc.edu/biblio/BRTF_Final_Report.pdf.

[210] With notable, isolated exceptions such as Clay Shirky, "Library of Congress Archive Ingest and Handling Test (AIHT) Final Report." National Digital Information Infrastructure & Preservation Program, June 2005, accessed April 9, 2020, http://www.digitalpreservation.gov/partners/aiht/high/ndiipp_aiht_final_report.pdf; and Brian Aitken et al., "The Planets Testbed: Science for Digital Preservation," *The Code4Lib Journal*, Issue 3 (2008-06-23). Unfortunately, both of these projects have concluded.

[211] As noted above, current certifications are based primarily on process rather than demonstration of efficacy or outcome conformance.

[212] For a possible approach see Christoph Becker and Andreas Rauber, "Decision Criteria in Digital Preservation: What to Measure and How," *Journal of the American Society for Information Science and Technology* 62.6 (2011): 1009-1028.

- Funders should give priority to programs that rigorously integrate research and practice.
- Research-based evaluations of practice should go beyond case studies in their approach, and include replicable methods to support systematic inference.

## 5.2 Stewardship at Scale

A cross-cutting research problem is dealing with challenges of scale in digital stewardship. These challenges are illustrated by two prominent preservation efforts: The End of Term Web Archive[213] and the termination of the Library of Congress's social media archive.[214] (See Sections 2.1, 2.2, and 2.2.2 for references to similar and related efforts).

The End of Term Web Archive, started in 2012, captures federal websites every four years, and exemplifies the successful growth of digital collections: in the most recent 2016-2017 round of harvesting, the collection expanded by over 50 percent (100TB) since its 2012 harvest with the inclusion of government-published databases and targeted social media collection. Not only is the content permanently archived by multiple institutions, the project provides access to the content as well. By 2018 the most recent data was made continuously publicly accessible online through the Internet Archive, and through its own portal, hosted by the California Digital Archive (see Section 2.1 for description of other related efforts).

By contrast, in 2017, the Library of Congress (LOC) discontinued its initiative to collect all Twitter posts.[215] This was based on a landmark agreement with Twitter in 2010 that permitted the Library to permanently archive all Twitter posts, and that provided an ingest workflow to LOC to support preservation. In part, this discontinuation was a result of the rapid expansion of the collection—the volume of Twitter posts grew by multiple orders of magnitude during this period. The discontinuation of the program illustrates the challenges of collection scaling—even for an institution as experienced and capable as LOC.

However, another critical challenge of the LOC Twitter archive was providing access to this collection. The amount of resources needed to provide indexing and searching across the collection dwarfed the resources needed for archiving (to underline the difficulties of access, the collection remains closed to the public indefinitely).  The scale of the collection creates a demand for summary analytics that supports collection-level analysis—which presents a new type of demand on the Library's collection. Further complicating the issue, the archive included tweets that were later deleted; offering access to the collection raised new privacy challenges—which would require a new access policy that balances the rights of the individual against the accuracy of the historical record. Moreover, the archiving agreement between Twitter and LOC remains unique—Twitter's terms of use effectively prevents systematic archiving of this collection by a different institution.[216]

---

[213] See for a description: Mark E. Phillips and Kristy K. Phillips, "End of Term 2016 Presidential Web Archive," *Against the Grain*, 29 (6) (December 2017-January 2018).

[214] Michael Zimmer, "The Twitter Archive at the Library of Congress: Challenges for Information Practice and Information Policy," *First Monday* 20, no. 7 (2015).; Kate Zwaard et al., "Institution as Social Media Collector: Lessons Learned from the Library of Congress." In *Proceedings of IFLA WLIC*, Kuala Lumpur, Malaysia, 2018, http://library.ifla.org/2428/1/093-zwaard-en.pdf.

[215] "Update on the Twitter Archive at the Library of Congress," Library of Congress, December 2017, accessed April 22, 2020, https://blogs.loc.gov/loc/files/2017/12/2017dec_twitter_white-paper.pdf.

Both the End of Term Web Archive and the LOC social media initiative exemplify how the changing scale and nature of collections affect selection and appraisal. Both efforts target collections of information that are of future interest to many distinct communities of research and practice, and both target (at least partially) publicly available information that is being selected and appraised by other organizations. Further, the End of Term Web Archive is designed as an explicitly collaborative effort: it is currently led by four prominent stewardship organizations (the Library of Congress, California Digital Library, Internet Archive, and the University of North Texas), that divide responsibilities for storage, access, collection development, and tool development. The project also engages in public dissemination and publicity so that a broad spectrum of curators (including librarians, archivists, researchers, and citizens) can participate in selecting content (e.g. through the public link proposal interface).

Moreover, effective scaling requires managing changes in multiple dimensions: changes in the scale of collections, changes in the scale of access, and changes in the scale of selection and appraisal.

Over the last five years, the scale of collections preserved by stewardship institutions continues to increase rapidly—and it is now not uncommon for collections to enter the range of petabytes.[217] Keeping track of everything and being able to work with and manage content is increasingly difficult. Growing volumes of digital materials test the financial and operational capabilities of organizations engaged in preservation activities. Of particular concern are issues around the stewardship of "big" data and the search and indexing of digital collections at scale.

---

[216] Michael Beurskens, "Legal Questions of Twitter Research." In *Twitter and Society*, Volume 89: 123-133, http://snurb.info/files/2014/Twitter%20and%20Society%20-%20Structural%20Layers%20of%20Communication%20on%20Twitter%20%282014%29.pdf.; Peter Lang, 2014.; David O'Brien et al., "Integrating Approaches to Privacy Across the Research Lifecycle: When Is Information Purely Public?" Berkman Center Research Publication No. 2015-7, posted March 29, 2015, accessed April 9, 2020, https://papers.ssrn.com/sol3/papers.cfm?abstract_id=2586158.

[217] Michelle Gallinger et al., "Trends in Digital Preservation Capacity and Practice: Results from the 2nd Bi-Annual National Digital Stewardship Alliance Storage Survey." *D-Lib Magazine*, Volume 23, Number 7/8 (July/August 2017), https://doi.org/10.1045/july2017-gallinger.

"Big" collections can create scaling challenges not only as a result of the sheer number of bits stored, but for other reasons, including the numbers of objects that must be curated, the velocity (frequency) with which data objects and collections are updated, and the variety (heterogeneity) of the data objects, formats, and characteristics. Thus scaling challenges go far beyond the bare provisioning of storage—with variety often being the biggest challenge for institutions.[218] Scaling to billions of files, and/or to individual files of extremely large size, renders manual methods of archival selection, quality evaluation and control all but impossible, creates performance challenges for data ingestion workflows and tools, increases the complexity of indexing and discovery, and may render standard computer-human interfaces used for curation and user access unusable.[219]

There is a well-identified taxonomy of potential risks to information loss. These risks include media failure, hardware failure, software failure, communication errors, network failure, media and hardware obsolescence, software obsolescence, operator error, natural disaster, external attack, internal attack, economic failure, and organizational failure.[220] Scaling collections presents special challenges for managing these risks: the increased number and size of files, and increased volume of collections, can overwhelm current approaches to replication, fixity checking, and repair that are needed to ensure long-term data integrity.[221] Increasing the number of formats and object types creates challenges for the in-depth documentation, format characterization, and format migration that are required to maintain long-term accessibility. Increasing amounts of data create particular challenges for maintaining the versioning and provenance required of durable, authentic collections.

Currently, many organizations lack the expertise or economies-of-scale to process and store big collections—which has led to greater reliance on cloud-storage providers for primary storage and/or replication.[222] While cloud storage is an appropriate part of a preservation strategy it is critical to diversify across providers and to regularly verify fixity across all replicas of content to manage collection risk. However, the practices of many cloud storage providers (see Section 4.1 for examples)—such as tiered egress charges, fixity caching, and subcontracting to third-party storage providers[223]—are opaque and promote lock-in. This creates substantial challenges to assessing and mitigating the preservation risk associated with these providers.

---

[218] Kevin C. De Souza, "Realizing the Promise of Big Data: Implementing Big Data Projects," 2014, http://www.businessofgovernment.org/sites/default/files/Realizing%20the%20Promise%20of%20Big%20Data.pdf.; Line Pouchard, "Revisiting the Data Lifecycle with Big Data Curation," *International Journal of Digital Curation*, 10(2) (2015), pp.176-192, http://www.ijdc.net/article/view/10.2.176.

[219] See e.g. Sara Day Thomson and William Kilbride, "Preserving Social Media: The Problem of Access." *New Review of Information Networking,* 20(1-2) (July 3, 2015): 261-75; Sara Day Thomson. "Preserving Transactional Data," May 2016, https://www.dpconline.org/docs/technology-watch-reports/1525-twr16-02/file.

[220] David S.H. Rosenthal et al., "Requirements for Digital Preservation Systems: A Bottom-Up Approach." *D-Lib Magazine*, Volume 11 Number 11 (November 2005), http://www.dlib.org/dlib/november05/rosenthal/11rosenthal.html.

[221] For general issues see David S.H. Rosenthal, "Bit Preservation: a Solved Problem?" *International Journal of Digital Curation*. 5.1 (2010): 134-148. For an examination of how selected collection types challenge current archival practices see: Sara Day Thomson and William Kilbride, "Preserving Social Media: The Problem of Access." *New Review of Information Networking*. 20(1-2) (July 3, 2015): 261-75.; Sara Day Thomson. "Preserving Transactional Data," May 2016, https://www.dpconline.org/docs/technology-watch-reports/1525-twr16-02/file.

As the scale of collections has grown, the scale of access has presented several types of challenges. First, the computational resources needed to offer standard access services (e.g. indexing and search) for a collection can increase dramatically with the size of the collections— as the Library of Congress's experience with Twitter demonstrates. Second, digitization substantially increases the number of people and the variety of communities using collections. This is a generally a positive result—but may put further strain on resources, create demand for new types of access (such as collection-level analytics), and probe the limits of existing policies.

A corollary of the increasing size of collections and breadth of access is that the collections may reveal answers to questions that were completely unanticipated at the time of acquisition. This has particularly large implications for privacy: neither the traditional methods of de-identification and anonymization nor the historical policies for managing confidentiality and personally-identifiable information are sufficient for protecting privacy across big-data collections.[224] While the privacy implications of big collections are now a focus of attention for the US Census and for some high-tech companies such as Google,[225]

---

[222] See Gallinger et. al 2017, above.

[223] NDSA Fixity Survey Working Group. "2017 Fixity Survey Report," 2018, https://osf.io/grfpa/; David S.H. Rosenthal and Daniel L. Vargas, "Distributed Digital Preservation in the Cloud," *International Journal of Digital Curation*. 8(1) (June 14 2013):107-19.; Kan Yang and Xiaohua Jia, "Data Storage Auditing Service in Cloud Computing: Challenges, Methods and Opportunities," *World Wide Web* 15 no. 4 (July 2011), 409-28.

[224] President's Council of Advisors on Science and Technology, "Big Data and Privacy: A Technological Perspective," 2014. https://bigdatawg.nist.gov/pdf/pcast_big_data_and_privacy_-_may_2014.pdf.; Micah Altman et al., "Practical approaches to big data privacy over time," *International Journal of Data Privacy Law* 8, issue 1 (February 2018): 29-51, https://doi.org/10.1093/idpl/ipx027; National Academies of Sciences, Engineering, and Medicine, "Federal statistics, multiple data sources, and privacy protection: Next steps," (Washington, D.C.: National Academies Press, 2017), https://doi.org/10.17226/24893.

[225] See Jerome P. Reiter, "Differential Privacy and Federal Data Releases," *Annual Review of Statistics and its Application* 6:85-101 (March 2019).; Kobbi Nissim et al., "Differential Privacy: A Primer for a Non-Technical Audience," *JETLaw* 21:209 (2018).

the implications for memory institutions are just starting to be explored.[226]

Finally, scaling the selection and appraisal process presents what is arguably the most fundamental challenge to traditional preservation practice and research. This is because the proliferation of digital materials qualitatively changes the availability of content for selection and access, and the marginal costs of additional replication and use—which in turn changes the opportunities and incentives for organizational coordination over the selection, curation, and use of content.

Because information produced in analog form is costly to replicate and to access non-locally, selection has historically been closely tied to organizational production and acquisition. What many organizations selected for preservation was naturally a subset of what was created by the organization or formally acquired for its operational use. Even those organizations that explicitly acquired external content for preservation purposes were substantially limited by their ability to obtain new content.

As processes such as mass digitization have lowered the cost of universal access, institutions have come to rely on large amounts of information that lie beyond their institutional boundaries. Modern selection policies may reasonably consider not only the information that an organization or designated community possesses, but what additional information that community uses, and what information would be valuable to it in the future. An institution may thus select from large portions of the web, social media, government documents, or research evidence base.[227]

At the same time, other institutions may be selecting from the same potential content—and with plans to make it publicly available. Thus a modern selection policy should seek to assess and identify to its community content that is both of value and at risk, given the selection strategies of other memory institutions. This creates a range of practical opportunities for coordination across stewardship organizations. However, our understanding of the reliability, design, and behavior of coordinated stewardship networks 2020 NDSA Agenda for Digital Stewardshipremains in its early stages.[228] Designing effective technical diversification strategies for long-term access requires more extensive modeling along these lines, and it creates a pressing need for research in economic, legal, and policy mechanisms to govern the knowledge commons.[229]

---

[226] See The New Media Consortium. "NMC Horizon Report: 2016 Museum Edition," 2016, https://library.educause.edu/~/media/files/library/2016/1/2016hrmuseumEN.pdf.; See also Zimmer 2015, above; Pekka Henttonen, "Privacy as an Archival Problem and a Solution," *Archival Science* 17.3 (2017): 285-303; Tara Robertson, "Not All Information Wants to be Free: The Case Study of On Our Backs." In *Applying Library Values to Emerging Technology: Decision-Making in the Age of Open Access, Maker Spaces, and the Ever-Changing Library*. Publications in Librarianship #72. American Library Association, 2018.

[227] On how digitization has affected curation and collective action see: Lorcan Dempsey, Constance Malpas, and Brian Lavoie, "Collection directions: The Evolution of Library Collections and Collecting," *portal: Libraries and the Academy* 14, 3 (2014): 393-423.; Micah Altman and Marguerite Avery, "Information Wants Someone Else to Pay For it: Laws of Information Economics and Scholarly Publishing," *Information Services & Use* 35, 1-2 (2015): 57-70.

Regardless of the effectiveness of coordination, appraisal is still necessary. The production of information far outstrips the collective capacity of stewardship organizations to select it and make it available. For example, the 40 petabytes of content now offered by the Internet Archive, one of the world's largest public archives, is a small fraction of the digital information that is being produced across the world every day.[230] It is neither desirable nor feasible to keep all information. Scaling selection requires scaling appraisal. Many current appraisal practices are based on labor-intensive expert judgments that rely on curators' deep expertise in understanding the needs of the organizations in which they are embedded and the communities they directly serve. These practices, judgments and expertise remain critical and are challenged by the vastly increasing size of collections, the breadth of content from which selections can be made, and the variety of communities which will access the content in the future. In essence, appraisal involves making predictions about the future use and value of potential collections by and to designated communities. Quantifying the future value of information is notoriously difficult, and in some cases impossible.[231] Thus, correctly estimating the future value of a single specific information object may be quixotic—similar to trying to guess the future stock price of a single corporation.

Curators continually make implicit expert judgments regarding what information to retain, how long to retain it, what effort to expend in making it accessible and understandable.

---

[228] See for example, Carly Dearborn and Sam Meister, "Failure as Process: Interrogating Disaster, Loss, and Recovery in Digital Preservation," *Alexandria: The Journal of National and International Library and Information Issues*, 27, 2 (2017): 83-93; Oya Y. Rieger *The State of Digital Preservation in 2018 A Snapshot of Challenges and Gaps*, Ithaka S+R. Last Modified 29 October 2018, accessed April 10, 2020, https://doi.org/10.18665/sr.310626.

[229] See Charlotte Hess and Elinor Ostrom, *Understanding Knowledge as a Commons: From Theory to Practice*, (Cambridge, MA: MIT Press, 2006).; Brett M. Frischmann, Michael J. Madison, and Katherine J. Strandburg, eds., *Governing Knowledge Commons* (Oxford University Press, 2014).; Micah Altman and Chris Bourg. "A Grand Challenges-Based Research Agenda for Scholarly Communication and Information Science," 2018, https://grandchallenges.pubpub.org/pub/final.

[230] See respectively, Martin Hilbert and Priscilla López, "The World's Technological Capacity to Store, Communicate, and Compute Information," *Science* 332, 6025 (2011): 60-65; and Nathan Mattise, "The Internet's Keepers? Some callus hoarders - I like to say we're archivists," October 7, 2018, accessed April 17, 2020, https://arstechnica.com/gaming/2018/10/the-internets-keepers-some-call-us-hoarders-i-like-to-say-were-archivists.

[231] Kenneth J. Arrow, "The Value of and Demand for Information." In *Decision and Organisation*. North-Holland, 1972.

Moreover, the size and scope of current collections requires that appraisal valuation be made more explicit, automatable, and empirical. Big-tech companies such as Google are focused on developing measures of 'value' that are scalable. The challenge is that current large-scale algorithms for evaluating information focus only on current relevance, broad appeal, and commercial potential (since the companies building these services run on ad revenue). As a recent analysis of grand challenges in information science highlights, algorithms such as those used by commercial vendors rely heavily on the monetary value that can be derived from such systems (such as sales of goods or ad placements). What is needed are empirically testable methods that can be used to estimate the future benefits to specific designated communities that will derive from access to the proposed collection.[232]

**Actionable Recommendations**

- Funders and researchers should prioritize programs and projects that increase the scalability of digital stewardship.
- Researchers should recognize that the challenges of "big" collections goes beyond size and storage, and includes dealing with the variety and velocity of big data and big collections across all phases of the curation lifecycle.
- Researchers and funders should recognize that selection and appraisal is a fundamental challenge at scale, and should prioritize systematic, evidence-based, non-labor intensive methods of evaluating portfolios of information.

## 5.3 Targeted Applied Research Areas

A number of research issues are less universal than those of scale and evidence, but are vital in order to develop more effective, reliable, and efficient tools, models, and methods for digital stewardship in the next three to five years.

**Actionable Recommendations**

- Funders and researchers should prioritize a number of targeted applied areas of research that constitute special opportunities for improving the reliability and efficiency of preservation practice, including: cost modeling, environmental sustainability, computability, and frameworks for trustworthiness at the level of the document, collection, and institution.

### 5.3.1 Applied Research for Cost Modeling

In the near term, there are specific areas of applied research around digital preservation lifecycle costs that need attention. These were called out in previous editions of the *Agenda*, and remain one of the top priorities (along with others such as advocating for preservation resources) across steward institutions in 2018. See Section [1.1](#) for a summary of survey results, and the [Appendix](#) for details.

---

[232] See Altman and Bourg (2018) above.

Currently, there are limited models for cost estimation for ongoing storage of digital content; cost estimation models need to be robust and flexible. There are bodies of written research on the topic that explore the costs of specific use cases.[233] However, overall, there has been little published over the last three years that substantially increases the discipline's ability to model curation costs in general. Simultaneously, at the level of practice, cloud storage options have made short- and medium-term cost models more challenging. For example, the structure of cloud storage fees and options have changed relatively rapidly, and fee structures vary non-linearly over factors such as number of replications, replication quality, egress and ingress velocity, and collection size.

Notwithstanding, there are significant limitations of stewardship cost models at both the abstract and applied level. Evidence from the NDSA staffing survey discussed in Section [3.2](#) suggests that there is a need to more clearly identify and characterize and the staffing aspects of digital stewardship, to understand the dissatisfaction among preservation staff, and to predict and plan for the resources necessary to "move from externally funded projects to scoped and well thought out internally supported programs." In addition, as discussed below in [5.3.2](#), there is a complementary need to develop high-level models that systematically and reliably predict the future value of preserved content—so that the cost-benefit ratio of different stewardship and collection models can be evaluated. Many long-term cost models are based on assumptions that the historical rate of decrease in storage prices will continue indefinitely—an assumption that is contradicted by a careful analysis of cloud storage trends and emerging storage technologies costs.[234] Furthermore, storage is only one component of long-term preservation costs, and the cost of curation processes such as media migration, format migration, and integrity checking are often missing from simple cost models. Moreover, media designed for long-term storage is unlikely to emerge as a cost-effective alternative.[235] These media are typically optimized for offline use, implying that durable media are a niche market in today's online world and are therefore unable to take advantage of economies of scale in production. Further, even where adopters are willing to pay for preservation, the up-front cost compares unfavorably to media migration with standard media, in scenarios where storage costs in general are expected to decline.

---

[233] See Luís Corujo, Carlos Guardado da Silva, and Jorge Revez. "Digital curation and costs: approaches and perceptions," In *Proceedings of the Fourth International Conference on Technological Ecosystems for Enhancing Multiculturality*. ACM, 2016.; and Butch Lazorchak, "A National Agenda Bibliography for Digital Asset Sustainability and Preservation Cost Modeling," January 14, 2014, accessed April 17, 2020, http://blogs.loc.gov/digitalpreservation/2014/01/a-national-agenda-bibliography-for-digital-asset-sustainability-and-preservation-cost-modeling/.

[234] David S.H. Rosenthal et al., "The Economics of Long-Term Digital Storage." In *Proceedings of Memory of the World in the Digital Age*, Vancouver, BC, 2012.

[235] David Stuart Holmes Rosenthal, "The Medium-Term Prospects for Long-Term Storage Systems," *Library Hi Tech* 35, no. 1 (2017): 11-31.

These complexities and the uncertainties of predicting costs and returns on stewarding collections create problems not only for individual organizations, but also for the development of shared services and common infrastructure. For example, these issues likely contributed to the recent shutdown of the Digital Preservation Network, discussed in detail in Section 4.1. Different approaches to cost estimation should be explored and compared to existing models with emphasis on reproducibility of results. The development of a cost calculator would benefit organizations in making estimates of the long-term storage costs for their digital content.

Further, as discussed in other sections, there are many opportunities to develop better value models and business models: in Stewardship at Scale (Section 5.2), we discuss the challenges of systematically and reliably predicting the future value of portfolios of preserved content. In Coordinating an Ecosystem of Sustainable Shared Services (Section 4.1), we discuss new collaboration and the need for new collaborative business models. A combination of value, cost, and business model development is needed for rational and efficient digital curation.This research needs to address multiple storage models: locally stored data, distributed preservation networks, data cooperatives, cloud storage, brokered cloud storage systems and hybrid systems should each be addressed in cost models so that organizations can make informed and cost-effective digital preservation decisions.

### 5.3.2 Environmental Sustainability and Sustainability of Digital Collections

As our digital cultural and scientific heritage grows at an exponential rate, it is often easy to overlook the underpinning material costs. Data, of course, are not "virtual" or "ephemeral"; rather, every byte requires resources to ensure its reliable storage and accessibility. Reports suggest that data management taxed upwards of 2% of total global energy consumption in 2012,[236] and the percentage has likely increased since then. There is a developing body of work on low-carbon "green" computing and data centers.[237] Metrics by which to evaluate the operational costs of data centers, such as those produced by the non-profit organization Green Grid, or the Jisc-funded Greening Information Management Assessment Framework,[238] offer ways in which digital preservationists may conduct preliminary, quantifiable assessments. However, there has been relatively little progress in developing and validating[239] these assessments. And there is still no substantive body of work connecting this work on environmental sustainability to economic modeling for long-term digital storage.

---

[236] James Glanz. "Data Centers Waste Vast Amounts of Energy, Belying Industry Image." *The New York Times*, September 22, 2012, http://www.nytimes.com/2012/09/23/technology/data-centers-waste-vast-amounts-of-energy-belying-industry-image.html.; Keith L. Pendergrass et al., "Toward Environmentally Sustainable Digital Preservation," *The American Archivist* 82 No. 1 (Spring/Summer 2019): 165-206.

[237] The Climate Group, "SMART 2020: Enabling the Low Carbon Economy in the Information Age," Global eSustainability Initiative, 2008.; Gary Cook, "How Clean Is Your Cloud?" Greenpeace, April 2012, http://www.greenpeace.org/international/Global/international/publications/climate/2012/iCoal/How

Exacerbating the environmental problems associated with digital infrastructure is the dramatic increase, over the last four years, of blockchain, which are used to provide the persistent, distributed ledgers on which cryptocurrencies and other digital infrastructure rely. As currently implemented (incorporating proof-of-work), blockchain systems are energy-intensive, and substantial media attention has been given to the exorbitant energy demands of these technologies and the need to regulate them.[240] Despite the environmental impacts of these solutions and the immaturity of institutional persistence models for them, blockchains are also increasingly proposed[241] as solutions to information persistence.

In contrast to "persistence" solutions such as blockchain, there are a number of approaches to information storage media that require marginal energy after writing, are potentially highly durable, and provide high information storage density.[242] While some of

---

CleanisYourCloud.pdf; "Google's Green Data Centers: Network POP Case Study," Google, accessed April 10, 2020, http://static.googleusercontent.com/external_content/untrusted_dlcp/www.google.com/en/us/corporate/datacenter/dc-best-practices-google.pdf.; Eric Masanet, Arman Shehabi, and Jonathan Koomey. "Characteristics of Low-carbon Data Centres," *Nature Climate Change* 3, no. 7 (July 2013): 627-630, doi:10.1038/nclimate1786.

[238] Diane McDonald, "Greening Information Management: a Focussed Literature and Activity Review," University of Strathclyde (2009): 1-42.

[239] Validation is a general challenge in the broader area—even with widely used and mature standards, such as those used in 'green' construction,see: Dat Tien Doan et al., "A Critical Comparison of Green Building Rating Systems," *Building and Environment* 123 (2017): 243-260.

[240] Jon Truby, "Decarbonizing Bitcoin: Law and Policy Choices for Reducing the Energy Consumption of Blockchain Technologies and Digital Currencies," *Energy Research & Social Science* 44 (2018): 399-410.

[241] Victoria L. Lemieux, "A Typology of Blockchain Recordkeeping Solutions and Some Reflections on Their Implications for the Future of Archival Preservation," In *2017 IEEE International Conference on Big Data*, IEEE, 2017 pp. 2271-2278.

[242] See, for a review, Melissa Guzman, Andreas M. Hein, and Chris Welch, "Extremely Long-Duration Storage Concepts for Space," *Acta Astronautica* 130 (2017): 128-136.

these technologies, such as DNA storage, are still in development,[243] others, such as silica-glass DVDs, are within the state of manufacturing practice.[244]

Thus, the issue of environmental sustainability should be seen as an economic problem as much as (or more than) a technical one. For sustainable long-term storage to be economically competitive, it needs to be produced at scale—however, most of the storage industry is driven by the need for cloud computing, which requires access latencies that are currently beyond the reach of these technologies, and which generally advantage energy-using, always-on storage. Further, in many countries, the price of electricity does not incorporate the full social costs of carbon and other emissions produced by electricity generation, putting green technologies at further disadvantage.

A comprehensive examination of the environmental sustainability of digital preservation requires an interdisciplinary perspective that merges material and access needs, and that brings together experts in digital preservation, information technology, computer science, environmental science and economics. These are only first steps, however, and a much more comprehensive, interdisciplinary approach is needed that takes into account issues of digital stewardship, and the economic externalities that blunt incentives for efficient durable storage. There is a need for basic research and development, in particular new case studies that could refine current metrics, as well as a need to investigate ways of educating the broader community about sustainability.

### 5.3.3 Research on Trust Frameworks

When describing systems formally, the term "trustworthiness" is often used to designate the degree to which that system can be expected to fulfill its designated functions, and the properties of the system that are causally related to this expectation. Less formally, systems are trustworthy when we have good reason to believe that they will work correctly. Trustworthiness is context-dependent—it may be defined with respect to different designated functions and for systems described at multiple levels—from the macro-level organization to micro-level of the technology.

Trustworthiness is thus a core requirement of successful digital preservation. An important goal of preservation research is to develop theoretically coherent and empirically reliable approaches to determine the trustworthiness of preservation organizations, of preservation-related services (especially cloud-based storage services), and of the information extracted from preserved collections.

---

[243] Substantial incremental progress is being made in this area, see, for example: Reinhard Heckel, "An Archive Written in DNA." *Nature Biotechnology* 36, no. 3 (2018): 236.

[244] See, for example, work underway at Microsoft: "Project Silica", accessed April 29, 2020, https://www.microsoft.com/en-us/research/project/project-silica/.

The importance of organization-level trustworthiness remains high. Much content has been lost to organizational failure.[245] Further, organization-level failure risks are part of a well-established taxonomy of risks any trustworthy preservation system should mitigate.[246] Moreover, while the last decade's advances in computing technology has increased the robustness of digital preservation systems in some areas, it has not increased (and likely decreased) organization robustness (see Section 3 and Section 4.1).

Individual stewardship organizations are often subject to a wide range of potential risk factors for catastrophic failure. These include changes in the local legal regime, catastrophic weather or war events in specific geographical areas, curatorial error, internal or external malfeasance, economic downturn, or change in organizational mission or leadership. Addressing these risks often require that content, the auditing of content, and the evaluation of organizations themselves be diversified across multiple organizations and stakeholders.[247]

In this area, community use of collaborative institutional mechanisms to mitigate preservation risk is growing. This is reflected in the growth of organizations such as the Global LOCKSS Network,[248] Data-PASS,[249] MetaArchive, the Digital Preservation Coalition, Chronopolis, CLOCKSS, and APTrust. These organizations, and the multi-institutional stewardship approach they represent, have increased both in use and in recognition—but have been subject to failures as well (see Section 4.1)

The preservation community has made considerable progress towards articulating the practices and behaviors of trustworthy preservation organizations, and in establishing some ways of documenting these standards and practices. At the more formal end of the spectrum, the ISO 16363 standard[250] enumerates elaborate sets of criteria for good practice, for conducting records-based audits of practice and organizational health, and for certifying auditors. The CoreTrustSeal (originally named *Data Seal of Approval)* provides a lighter-weight mechanism for certifying repository practice.[251] On the other end of the spectrum of complexity—the NDSA *Levels of Preservation*[252] provides a concise, technically-focused inventory of criteria that are believed to increase the trustworthiness of organizations.

---

[245] As an illustrative example, see Wikipedia's list of destroyed libraries and archives: https://en.wikipedia.org/wiki/List_of_destroyed_libraries, which documents some of the most dramatic cases of organizational failure.

[246] David S.H. Rosenthal et al., "Requirements for Digital Preservation Systems: A Bottom-Up Approach." *D-Lib Magazine*, Volume 11 Number 11 (November 2005), http://www.dlib.org/dlib/november05/rosenthal/11rosenthal.html. HYPERLINK "about:blank" Micah Altman, Bryan Beecher, and Jonathan Crabtree, "A Prototype Platform for Policy-Based Archival Replication," *Against the Grain* 21(2) (2009): 44-47, http://www.data-pass.org/sites/default/files/ATGpre.pdf.; Micah Altman and Jonathan Crabtree, "Using the SafeArchive System: TRAC-Based Auditing of LOCKSS." In *Proceedings of Archiving 2011*. Society for Imaging Science and Technology, 2011.

[248] "Global LOCKSS Network," LOCKSS, accessed April 16, 2020, https://www.lockss.org/join-lockss/networks/global-lockss-network.

[249] "Data-PASS," Data Preservation Alliance for the Social Sciences, accessed April 16, 2020, http://www.data-pass.org/.

These initiatives notwithstanding, the reliable evaluation of the trustworthiness of preservation organizations, services remains a substantial challenge for policy research for three reasons. First, memory institutions have been slow to adopt these mechanisms and related practices have not received general recognition or adoption[253]—with the notable exception of the adoption of the CoreTrustSeal[254] by over 60 (primarily European) repositories between August 2017 and July 2019. Second, current trusted repository approach utilizes a very limited subset of the available mechanisms generally employed in trust engineering[255]—primarily records-based auditing, and self-review. Third, the reliability, effectiveness, and costs of current trust frameworks has yet to be systematically empirically demonstrated and systematically measured.

In general, further research is needed in the design, implementation, and evaluation of trustworthy digital stewardship mechanisms and their use including: building an organization's capacity to demonstrate trustworthiness, rewards, and penalties; peer review; statistical quality control and reliability estimation; incentive compatible mechanisms; threat-modeling and vulnerability assessment; portfolio diversification models; transparency and the release of information permitting direct evaluation of compliance; cryptographic approaches, including cryptographic signatures over semantic content; and generating and managing social evidence of compliance.

---

[250] International Standards Organization. *Space data and information transfer systems—Audit and certification of trustworthy digital repositories*. ISO 16363:2012. Geneva, Switzerland: ISO, reviewed and confirmed 2017.

[251] Mary Vardigan and Jared Lyle, "The Inter-University Consortium for Political and Social Research and the Data Seal of Approval: Accreditation Experiences, Challenges, and Opportunities," *Data Science Journal* 13 (2014):PDA83-7.

[252] Megan P. Phillips et al., "The NDSA Levels of Digital Preservation: Explanation and Uses." In *Archiving Conference*, Vol. 2013, No. 1, pp. 216-222. Society for Imaging Science and Technology, 2013, and the resources found on the OSF site for the "2019 Levels of Digital Preservation," 2019, DOI [10.17605/OSF.IO/QGZ98](10.17605/OSF.IO/QGZ98).

[253] See Michelle Gallinger et al., "Trends in Digital Preservation Capacity and Practice: Results from the 2nd Bi-Annual National Digital Stewardship Alliance Storage Survey." *D-Lib Magazine*, Volume 23, Number 7/8 (July/August 2017), https://doi.org/10.1045/july2017-gallinger.

[254] Core Trust Seal, "Core Certified Repositories," accessed April 28, 2020, https://www.coretrustseal.org/why-certification/certified-repositories/.

[255] See Bruce Schneier, *Liars and Outliers*, John Wiley & Sons 2012 for a review of trust engineering approaches.

All of the limitations with respect to establishing organizational trustworthiness apply not only to self-described preservation organizations but also to organizations upon which stewards are relying. For example, when an organization relies solely on a cloud storage service such as Amazon Web Services to ensure the long-term integrity and accessibility of their data, they are for all intents and purposes trusting that organization for preservation. And in general, there is little evidence that commercial storage platforms should be trusted for this purpose. More often, preservation organizations make use of third-party services as part of a storage strategy, without fully delegating preservation trust. This is often accomplished through a combination of fixity and auditing practices.[256] In theory, such auditing can be done in a resource-efficient way through the use of appropriate cryptographic protocols.[257] However, the APIs exposed by cloud vendors do not currently provide ways of reliably verifying the integrity of content stored in the system—besides requesting a copy, and verifying that copy in one's own secure environment. For example, APTrust currently uses random "fire drills" as a way to test both system and depositor durability.[258] However, applied research is still needed in methods to enable efficient remote cryptographic verification of content stored in these services.

Enthusiasm for blockchain has extended to the publication and preservation of information, particularly in academia—and some claim that blockchain is a solution to trustworthy content.[259] More accurately, there is no single blockchain—but multiple competing implementations combining different sets of methods developed over decades of computer theory that aim to provide the affordance of a tamper-resistant, distributed, persistent, append-only, public ledger.[260] However, the use of these technologies for preservation and stewardship remains speculative, and successful use of these technologies for digital preservation would require addressing some practical and theoretical hurdles. In practice, blockchain remains highly vulnerable to bugs and attack in their implementations and at end-points.[261] Further, blockchain systems currently do not scale to operations over large volumes of data (they are typically used as ledgers for content hashes and signature), making them more attractive for integration in a distributed file system such as IPFS than as a direct solution for permanent collection storage.[262]

---

[256] For a discussion of the use of fixity and auditing to enforce a higher-level preservation policy, see Micah Altman and Jonathan Crabtree, "Using the SafeArchive System: TRAC-Based Auditing of LOCKSS." In *Proceedings of Archiving 2011*. Society for Imaging Science and Technology, 2011.

[257] See, for example James Hendricks, Gregory R. Ganger, and Michael K. Reiter, "Verifying Distributed Erasure-Coded Data." In *Proceedings of the Twenty-sixth Annual ACM Symposium on Principles of Distributed Computing* 139-146. ACM, 2007.

[258] Bradley Daigle, "Not your Childhood Fire Drills," DPC Blog, last updated June 27, 2019, accessed April 10, 2020, https://www.dpconline.org/blog/fire-drills.

[259] Joris van Rossum, *Blockchain for Research*. Digital Science, November 2017, https://www.digital-science.com/resources/digital-research-reports/blockchain-for-research/.

[260] Joseph Bonneau et al., "SoK: Research Perspectives and Challenges for Bitcoin and Cryptocurrencies. In *2015 IEEE Symposium on Security and Privacy*. IEEE, 2015.; Karl Wüst and Arthur Gervais, "Do you need a Blockchain?" In *2018 Crypto Valley Conference on Blockchain Technology (CVCBT)*. IEEE, 2018.

 A major theoretical hurdle for any distributed system (whether blockchain or file system) is that these systems achieve their theoretical properties only with a sufficient number of stakeholders actively and continuously participating (I.e. multi-institutional coordination matters—see Section 4.1). Current blockchain systems do not provide credible sustainable incentives for individuals and institutions to participate over long periods. In fact, systems like Bitcoin that rely on "proof-of-work" concepts have, by design, long-term disincentives (e.g. mining bitcoin, a major incentive for current participation in the system, becomes unprofitable over time). Other disincentives are poor scalability[263] and negative environmental impacts (see section 5.3.2).

When collections are not lost altogether due to organizational or technical failure, it remains important to establish trust in the preserved materials. Digital preservation succeeds to the extent that it enables communication with the future.[264] Trustworthiness of that essential communication is threatened when the content or its meaning can be altered without detection and a permanent record of that fact; when intentional transformations (e.g. to migrate file formats) introduce unintended changes in meaning; or when the meaning of the object can no longer be reliably understood because external information (format specification, contextual information) is no longer available or reliably known.

Because long-term management of digital content often involves changing the representation of that content while retaining its semantics, the concept of "significant properties"[265] of content—identifying the properties of that content that give it meaning— has emerged as a key concept in digital preservation, and generated a focused and influential body of research.[266] The concept of significant properties can be applied to all content types—and recent expansions focus on priority content areas such as software and data,[267] that are discussed in the Content section above (Section 2.2). A more qualitative approach focusing on the same problem is describing "preservation intent"[268] and providing evidence that it has been successfully accomplished. Moreover, this line of research has implications across a diverse set of applications including format selection and migration; quality measurement and control; rights management; and information discovery and retrieval.

---

[261] David Gerard, *Attack of the 50 Foot Blockchain: Bitcoin, Blockchain, Ethereum & Smart Contracts* (London: David Gerard, 2017).
[262] Yongle Chen et al., "An Improved P2P File System Scheme Based on IPFS and Blockchain," In *2017 IEEE International Conference on Big Data* 2652-2657. IEEE, 2017.
[263] Arvind Narayanan et al., *Bitcoin and Cryptocurrency Technologies: A Comprehensive Introduction* (Princeton, N.J.: Princeton University Press, 2016).
[264] Reagan Moore, "Towards a Theory of Digital Preservation," *International Journal of Digital Curation* 3, no. 1 (2008).
[265] This term was first coined in: Margaret Hedstrom and Christopher A. Lee, "Significant Properties of Digital Objects: Definitions, Applications, Implications." In *Proceedings of the DLM-Forum*, 218-27. 2002.

Although widely used in the commercial sector,[269] methods for scalable evaluation of semantic similarity is far less common in digital preservation practice. Yet, the multiplicity of instantiations of the same or similar digital objects illustrates the need for and application of basic research to explore the many ways multiple digital objects could contain equivalent informational content given different contexts of significance.

Preservation research needs to map out the networks of similarity and equivalence across different instantiations of objects so that they can make better decisions on how to manage content, bearing in mind what properties of a given set of digital objects are significant[270] to their particular community of use. Research is also required in order to characterize quality and fidelity dimensions and create methods for computing format-independent fingerprints of content,[271] so that the fidelity of digital objects can be effectively managed over time. Beyond basic research to develop methods for identifying information equivalence, there is a need for research in different usage contexts to understand when particular modes or levels of information equivalence are relevant to particular stakeholders in particular contexts.

Further establishing trustworthiness for dynamic, customized, and/or personalized content is difficult even with trustworthy archival organizations, storage, and timestamp information.[272] Even when the archived content itself is trustworthy, the meaning can be altered if the context (e.g. linked resources) for that content is not also protected and validated.[273] Given the ubiquity of content that is embedded, dynamic, and personalized, research in validating the semantics of content is even more vital.

---

[266] David Giaretta, *Advanced Digital Preservation* (Berlin: Springer-Verlag, 2011).

[267] Simone Sacchi et al., "A Framework for Applying the Concept of Significant Properties to Datasets." *Proceedings of the 74th Annual Meeting of the American Society for Information in Science and Scholarship*. 2011.

[268] See for a discussion, Trevor Owens, *The Theory and Craft of Digital Preservation* (Baltimore: Johns Hopkins University Press, 2018).

[269] See for example, this analysis of semantic similarity methods as used in commercial copyright enforcement: Emanuele Lunadei, Christian Valdivia Torres, and Erik Cambria. "Collective Copyright: Enabling the Natural Evolution of Content Creation in the Web Era." In *Proceedings of the 23rd International Conference on World Wide Web*. ACM 2014.

[270] Margaret Hedstrom and Christopher A. Lee, "Significant Properties of Digital Objects: Definitions, Applications, Implications." In *Proceedings of the DLM-Forum*, 218-27. 2002.

[271] Micah Altman, "A Fingerprint Method for Scientific Data Verification." In *Advances in Computer and Information Sciences and Engineering*, 311-316. Springer Netherlands, 2008.

# ACKNOWLEDGEMENTS

## About the Authors of the 2019 Edition

The NDSA Agenda Working Group authored this report and engaged in discussions to identify significant trends and challenges since the 2015 *Agenda*. The membership of the NDSA contributed to these discussions through surveys, review comments, and interviews. This dialog was enriched by an extensive range of resources and current research. The NDSA Coordinating Committee provided commentary and review.

**Micah Altman**
Dr. Micah Altman is Director of Research for the Center for Research on Equitable and Open Scholarship (CREOS), and Head/Scientist, Program on Information Science for the MIT Libraries, at the Massachusetts Institute of Technology. He previously served as a Non-Resident Senior Fellow at The Brookings Institution, and at Harvard University as the Associate Director of the Harvard-MIT Data Center, Archival Director of the Henry A. Murray Archive, and Senior Research Scientist in the Institute for Quantitative Social Science*s.*

**Bradley Daigle**
Bradley Daigle is content and strategic expert for the Academic Preservation Trust and other external partnerships at the University of Virginia Library. He also works on copyright issues related to digital collections. Currently he is also Chair of the Virginia Heritage Governance Team, Lead for the DPLA Service Hub *Digital Virginias*, and Chair of the NDSA Leadership Team. Having been in the library profession for over fifteen years, he has published and presented on a wide range of topics including mass digitization, digital curation and stewardship, sustaining digital scholarship, intellectual property issues, mentoring in libraries, and digital preservation. In addition to his professional field, his research interests also include the history of the book, natural history, and early modern British literature. He received his MA in literature from the University of Montréal and an MLS from Catholic University.

---

[272] See Mohamed Aturban, Michael L. Nelson, and Michele C. Weigle, "Difficulties of Timestamping Archived Web Pages," December 8, 2017, https://arxiv.org/pdf/1712.03140.pdf.
[273] Ada Lerner, Tadayoshi Kohno, and Franziska Roesner. "Rewriting History: Changing the Archived Web from the Present." In *Proceedings of the 2017 ACM SIGSAC Conference on Computer and Communications Security*, 1741-1755. ACM 2017.

**Karen Cariani**
Karen Cariani is The David O. Ives Executive Director of the WGBH Media Library and Archives (MLA) and WGBH Project Director for the American Archive of Public Broadcasting (AAPB). The MLA provides access to the WGBH archives by providing circulation, accessioning, and preservation activities, in addition to licensing services. Cariani has 30-plus years of television production and project management experience. She has been project director for numerous projects including: WGBH's Teachers' Domain, now PBS Learning Media; WGBH Open Vault; the Boston Local TV News Digital Library project and for development of a digital media preservation system utilizing the Hydra/Samvera technology in partnership with Indiana University. She served two terms (2001-2005) on the Board of Directors of Association of Moving Image Archivists (AMIA). She was co-chair of the AMIA Local Television Task Force, and Project Director of the guidebook "Local Television: A Guide To Saving Our Heritage," (funded by the NHPRC) and co-chair of the AMIA Copyright and AMIA Open Source Committees. She was co-chair of the NDSA Infrastructure Interest Group, and served as president of Digital Commonwealth. Recent projects include serving as WGBH Project Director for the American Archive for Public Broadcasting in partnership with the Library of Congress. She is active in the archive community and professional organizations and passionate about the use of media archives and digital library collections for learning and education, but has a particular affinity for science.

**Christie Moffatt**
Christie Moffatt is a member of the federal staff of the National Library of Medicine (NLM), National Institutes of Health, serving as the Program Manager of the Digital Manuscripts Program in the NLM History of Medicine Division and Chair of the NLM Web Collecting and Archiving Working Group. In these roles, she is involved with digitization and access to 20th century historical collections in science, medicine, and public health, and web archiving on topics and events related to NLM collecting interests, including Global Health Events (Ebola outbreak, Zika Virus, Novel Coronavirus), HIV/AIDS, and the opioid epidemic. Moffatt earned her MLIS from UNC Chapel Hill, and a BA in History from the University of Georgia.

**Sibyl Schaefer**
Sibyl Schaefer is the Chronopolis Program Manager and Digital Preservation Analyst for Research Data Curation at the University of California, San Diego. In addition to working with national digital preservation efforts like the NDSA, she helps define long-term digital preservation solutions for the UCSD campus. She previously served as the Head of Digital Programs for the Rockefeller Archive Center, where she worked to fully integrate digital and traditional archival practices, including policy development, forensic and accessioning workflows, and training initiatives to support the long-term stewardship of digitized and born digital materials. She has been recognized as an Emerging Leader by the American

Library Association and has participated in the Archival Leadership Institute. Schaefer holds an MLIS with a specialization in Archival Studies from the University of California, Los Angeles.

**Bethany Scott**
Bethany Scott is Digital Projects Coordinator at the University of Houston Libraries' Special Collections, where she is responsible for planning and coordinating the digitization of archival materials, implementing Archivematica and ArchivesSpace, and overseeing the Libraries' digital preservation program. She holds an MS in Information Studies from the University of Texas at Austin and a BA in Studio Art from the University of St. Thomas and the Glassell School of Art.

**Lauren Work**
Lauren Work is the Digital Preservation Librarian at the University of Virginia, where she is responsible for the implementation of preservation strategy and systems for university digital resources. She helps to create workflows and strategies for the sustainable ingest, preservation and access to born-digital content at Virginia, in collaboration within communities such as the Academic Preservation Trust, Archivematica, and BitCurator. She earned her Master of Library and Information Science degree from the University of Washington.

## Authors of Prior Editions

This work revises and extends the 2015 edition of the *Agenda*. The authors of the 2015 edition of the *Agenda* edition included:

Micah Altman (MIT), Jefferson Bailey (Internet Archive), Karen Cariani (WGBH), Jim Corridan (Indiana Commission on Public Records), Jonathan Crabtree (UNC, Chapel Hill), Michelle Gallinger (Library of Congress), Andrea Goethals (Harvard Library), Abbie Grotke (Library of Congress), Cathy Hartman (University of North Texas), Butch Lazorchak (Library of Congress), Jane Mandelbaum (Library of Congress), Carol Minton Morris (DuraSpace), Kate Murray (Library of Congress), Trevor Owens (Library of Congress), Megan Phillips (NARA), Abigail Potter (Library of Congress), Robin Ruggaber (University of Virginia), John Spencer (BMS/Chace), Helen Tibbo (UNC Chapel Hill), Kate Wittenberg (Portico).

The NDSA would also like to thank the reviewers and initial readers of the 2015 *Agenda* for their thoughtful comments, which greatly improved the document.

# APPENDIX

## 2018 NDSA Institutional Survey on Priority Areas

In order to revalidate the priorities called out in prior agendas, the NDSA Agenda Working Group created an integrated list of candidate challenges, drawn from past agenda documents, and other key reviews of the discipline. The Working Group then developed and disseminated a survey to gather information regarding these priorities and to probe for emerging concerns and initiatives.

Respondents to the survey were asked to provide information in the following areas:

- Rank the importance of the integrated list of candidate challenges to the discipline.
- Rank the challenges to their own organization—and describe organizational planning to meet those challenges.
- Identify new initiatives, projects and concerns.

In early 2018, the survey was sent to all NDSA institutional contacts. Eighty institutions completed the survey, which included a recontact of a subsample of nonresponders to probe for non-response bias. The final response rate was 38 percent, and there was no significant difference between initial responders and recontacted initial nonresponders. Telephone interviews were conducted with a 10 percent subsample of the responders to further probe for new initiatives, projects and concerns.

In aggregate, respondents identified the most important challenges to the preservation community as: (see Table A-1 for detailed results)

- Identifying and evaluating effective digital preservation practices
- Advocating for resources to support digital stewardship programs and activities
- Developing effective cost models for digital stewardship
- Maintaining ongoing integration, interoperability, and collaborative projects

2020 NDSA Agenda for Digital Stewardship81

**Table A-1 Ranking of Community Challenges**

|  | Ranking: (1) Very Unimportant … (5) Very Important | | | | | |
|---|---|---|---|---|---|---|
|  | 1 | 2 | 3 | 4 | 5 | Weighted Score |
| **Question** |  |  |  |  |  |  |
| Identifying and evaluating effective digital preservation practices | 0 | 3 | 6 | 32 | 42 | 338 |
| Advocating for resources to support digital stewardship programs and activities | 1 | 3 | 11 | 31 | 37 | 309 |
| Developing effective cost models for digital stewardship | 3 | 4 | 7 | 33 | 36 | 312 |
| Maintaining ongoing integration, interoperability, and collaborative projects | 0 | 1 | 11 | 36 | 35 | 319 |
| For your institution's mission, how important is it to address the following challenges?-Coordinating digital stewardship at scale | 3 | 1 | 15 | 32 | 29 | 273 |
| Privacy and ethical concerns | 1 | 2 | 12 | 37 | 29 | 293 |
| Integrating digital stewardship practice and thinking across an entire organization | 2 | 3 | 13 | 36 | 28 | 284 |
| Coordinating an ecosystem of sustainable shared stewardship services | 3 | 5 | 16 | 32 | 26 | 258 |
| Addressing urgent needs of specific types of at-risk digital content | 1 | 3 | 12 | 39 | 26 | 286 |
| Developing effective multi-institutional collaboration for digital stewardship | 2 | 7 | 13 | 35 | 25 | 265 |
| Appraising or forecasting the long-term use/value of digital collections | 0 | 5 | 14 | 37 | 25 | 273 |
| Engagement with content creators to leverage their incentives to preserve | 1 | 11 | 7 | 39 | 23 | 271 |
| Identifying the long-term risks to shared digital content | 2 | 6 | 14 | 39 | 22 | 266 |
| Accessibility of digital services and resources for those with disabilities | 2 | 3 | 18 | 37 | 22 | 258 |
| Economic and political pressures | 0 | 5 | 19 | 37 | 22 | 258 |
| Adapting organizational designs to the future of work | 1 | 7 | 20 | 35 | 19 | 235 |
| Embracing the need for radical organizational change | 3 | 11 | 25 | 25 | 18 | 190 |
| Supporting emerging digital stewardship professionals | 4 | 8 | 17 | 38 | 16 | 232 |
| Measuring the impact of new technologies | 1 | 8 | 17 | 42 | 14 | 238 |
| Evaluating community efforts to protect shared digital content | 4 | 7 | 26 | 34 | 12 | 196 |
| Integrating rigorous research and preservation practice | 5 | 2 | 19 | 47 | 10 | 238 |

Further, respondents indicated that the following two challenges were of high priority to their institutions: (see Table A-2 for detailed results)

- Integrating digital stewardship practice and thinking across an entire organization
- Economic and political pressures

## Table A-2. Organizational Priorities

|  | Ranking: (1) Very Unimportant … (5) Very Important | | | | | |
|---|---|---|---|---|---|---|
|  | 1 | 2 | 3 | 4 | 5 | Weighted Score |
| **Question** | | | | | | |
| Developing effective cost models for digital stewardship | 3 | 4 | 7 | 33 | 36 | 312 |
| Maintaining ongoing integration, interoperability, and collaborative projects | 0 | 1 | 11 | 36 | 35 | 319 |
| For your institution's mission, how important is it to address the following challenges?-Coordinating digital stewardship at scale | 3 | 1 | 15 | 32 | 29 | 273 |
| Privacy and ethical concerns | 1 | 2 | 12 | 37 | 29 | 293 |
| Integrating digital stewardship practice and thinking across an entire organization | 2 | 3 | 13 | 36 | 28 | 284 |
| Coordinating an ecosystem of sustainable shared stewardship services | 3 | 5 | 16 | 32 | 26 | 258 |
| Addressing urgent needs of specific types of at-risk digital content | 1 | 3 | 12 | 39 | 26 | 286 |
| Developing effective multi-institutional collaboration for digital stewardship | 2 | 7 | 13 | 35 | 25 | 265 |
| Appraising or forecasting the long-term use/value of digital collections | 0 | 5 | 14 | 37 | 25 | 273 |
| Engagement with content creators to leverage their incentives to preserve | 1 | 11 | 7 | 39 | 23 | 271 |
| Identifying the long-term risks to shared digital content | 2 | 6 | 14 | 39 | 22 | 266 |
| Accessibility of digital services and resources for those with disabilities | 2 | 3 | 18 | 37 | 22 | 258 |
| Economic and political pressures | 0 | 5 | 19 | 37 | 22 | 258 |
| Adapting organizational designs to the future of work | 1 | 7 | 20 | 35 | 19 | 235 |
| Embracing the need for radical organizational change | 3 | 11 | 25 | 25 | 18 | 190 |
| Supporting emerging digital stewardship professionals | 4 | 8 | 17 | 38 | 16 | 232 |
| Measuring the impact of new technologies | 1 | 8 | 17 | 42 | 14 | 238 |
| Evaluating community efforts to protect shared digital content | 4 | 7 | 26 | 34 | 12 | 196 |
| Integrating rigorous research and preservation practice | 5 | 2 | 19 | 47 | 10 | 238 |

Further, although respondents consistently identified many of these challenges as important to their organizations, less than six percent of respondents indicated that their organizations had a plan that addressed all of its important preservation challenges, and over 25 percent of organizations reported having no documented plans to address the most important preservation challenges for that organization (see Table A-3).

## Table A-3. Organizational preparedness

| Level of Organizational Preparedness | Proportion of Respondents |
|---|---|
| My organization has a plan that is likely to resolve the most important challenges. | 0.05814 |
| My organization has a plan that is likely to resolve some important challenges. | 0.50000 |
| My organization has a plan that is very likely to resolve some important challenges. | 0.11628 |
| My organization has no documented plan in these areas. | 0.25581 |
| Other (generally described nominal/weak plans) | 0.06977 |

The survey responses also identified a number of initiatives and tools in the preservation space, all of which were evaluated and incorporated into appropriate sections of the agenda document.